\documentclass[prb,amsmath,amssymb,superscriptaddress,twocolumn]{revtex4}
\usepackage{float}
\usepackage{amsmath}

\usepackage{amssymb}
\usepackage{amsfonts}
\usepackage{euscript}
\usepackage{enumerate}
\usepackage{hhline}
\usepackage{pslatex}
\usepackage{tabularx}
\usepackage[usenames,dvipsnames]{xcolor}

\usepackage{graphicx}% Include figure files
\usepackage{dcolumn}% Align table columns on decimal point
\usepackage{bm}% bold math
\usepackage[sort&compress]{natbib}

\makeatletter
\renewcommand{\@biblabel}[1]{#1. }
\renewcommand{\@dotsep}{500}
\renewcommand{\@pnumwidth}{0em}
\renewcommand{\l@figure}[2]{% #1 is e.g., Figure 1 + caption, #2 is pg.
\@dottedtocline{1}{1.5em}{2em}{Figure #1}{}\vspace{15pt}}

\definecolor{kartikcolor}{RGB}{166, 26, 20}

\definecolor{ARcolor}{RGB}{90,90,160}

\begin{document}

\title{Integrated photonic interposers for processing octave-spanning microresonator frequency combs}

\author{Ashutosh Rao}\email{ashutosh.rao@nist.gov}
\affiliation{Microsystems and Nanotechnology Division, Physical Measurement Laboratory, National Institute of Standards and Technology, Gaithersburg, MD 20899, USA}\affiliation{Maryland NanoCenter, University of Maryland,
College Park, MD 20742, USA}

\author{Gregory Moille}
\affiliation{Microsystems and Nanotechnology Division, Physical Measurement Laboratory, National Institute of Standards and Technology, Gaithersburg, MD 20899, USA}\affiliation{Joint Quantum Institute, NIST/University of Maryland, College Park, MD 20742, USA}
\author{Xiyuan Lu}
\affiliation{Microsystems and Nanotechnology Division, Physical Measurement Laboratory, National Institute of Standards and Technology, Gaithersburg, MD 20899, USA}\affiliation{Maryland NanoCenter, University of Maryland,
College Park, MD 20742, USA}
\author{Daron A. Westly}
\affiliation{Microsystems and Nanotechnology Division, Physical Measurement Laboratory, National Institute of Standards and Technology, Gaithersburg, MD 20899, USA}

\author{Davide Sacchetto}
\affiliation{Ligentec, EPFL Innovation Park, Bâtiment C, Lausanne, Switzerland}
\author{Michael Geiselmann}
\affiliation{Ligentec, EPFL Innovation Park, Bâtiment C, Lausanne, Switzerland}
\author{Michael Zervas}
\affiliation{Ligentec, EPFL Innovation Park, Bâtiment C, Lausanne, Switzerland}

\author{Scott B. Papp}
\affiliation{Time and Frequency Division, Physical Measurement Laboratory, National Institute of Standards and Technology, Boulder, CO 80305, USA}
\affiliation{Department of Physics, University of Colorado, Boulder, CO 80309, USA}
\author{John Bowers}
\affiliation{Department of Electrical and Computer Engineering, University of California, Santa Barbara, CA 93106, USA}

\author{Kartik Srinivasan} \email{kartik.srinivasan@nist.gov}
\affiliation{Microsystems and Nanotechnology Division, Physical Measurement Laboratory, National Institute of Standards and Technology, Gaithersburg, MD 20899, USA}\affiliation{Joint Quantum Institute, NIST/University of Maryland, College Park, MD 20742, USA}

\date{\today}% It is always \today, today,

\begin{abstract}
\noindent \textbf{Microcombs - optical frequency combs generated in microresonators - have advanced tremendously in the last decade, and are advantageous for applications in frequency metrology, navigation, spectroscopy, telecommunications, and microwave photonics. Crucially, microcombs offer the prospect of fully integrated miniaturized optical systems with unprecedented reductions in cost, size, weight, and power. However, this goal has been consistently hindered by the use of bulk free-space and fiber-optic components to process microcombs, limiting form factors to the table-top. Here, we address this challenge by introducing an integrated photonics interposer architecture to process microcombs and replace discrete components. Taking microcomb-based optical frequency synthesis in the telecom $C$-band around 1550~nm as our target application, we develop an interposer architecture that collects, routes, and interfaces octave-wide optical signals between photonic chiplets and heterogeneously integrated devices that constitute the synthesizer. We have implemented the octave spanning spectral filtering of a microcomb, central to the interposer, in the popular silicon nitride photonic platform, and have confirmed the requisite performance of the individual elements of the interposer. Moreover, we show that the thick silicon nitride needed for bright dissipative Kerr soliton generation can be integrated with the comparatively thin silicon nitride interposer layer through octave-bandwidth adiabatic evanescent coupling, indicating a path towards future system-level consolidation. Our interposer architecture addresses the immediate need for on-chip microcomb processing to successfully miniaturize microcomb systems. As microcombs and integrated devices evolve, our approach can be readily adapted to other metrology-grade applications based on optical atomic clocks and high-precision navigation and spectroscopy.} 
\end{abstract}

% \pacs{78.67.Hc, 42.70.Qs, 42.60.Da} \maketitle

\maketitle

Optical microcombs, generated in micro and nanophotonic resonators, have substantially broadened the reach of applications of optical frequency combs~\cite{Cundiff2003}. Along with the promise of a dramatic transformation from traditional table-top and rack-mount form factors to chip-scale integrated systems, a variety of applications have been shown to benefit from the use of microcombs~\cite{Kippenberg2018a,Gaeta2019,Pasquazi}. Furthermore, persistent innovation enabled by the precision nanofabrication of nanophotonic resonators continues to yield desirable and exotic optical microcombs~\cite{Moille2018,Yang2020,Zhao2020,Yu2020,Tikan2020} for next-generation systems. The convergence of nanophotonic resonators with scalable integrated photonics inherently supports the promise of creating integrated microcomb-based systems, with immediate applications in optical frequency synthesis~\cite{Jones2000,Holzwarth2000a,Spencer2018}, optical atomic clocks~\cite{Diddams2001,Newman2019}, optical distance ranging~\cite{Swann2006,Suh2018,Riemensberger2020}, optical spectroscopy~\cite{Thorpe2006,Suh2016,Dutt2018},  microwave and radiofrequency photonics~\cite{Fortier2011,Wu2018,Lucas2020}, astronomy~\cite{Li2008,Metcalf2019}, and telecommunications~\cite{Marin-Palomo2017,Fulop2018,Corcoran2020}.

However, to realize these integrated microcomb-based systems, integrated photonic interposers that connect and operate on optical signals that transit between the many constituent photonic components will be critical. In fact, the pursuit of such integrated systems has driven recent progress in active photonics, e.g., lasers~\cite{Li2018,Shtyrkova2019,Huang2019,Bhardwaj2020} and detectors~\cite{Yu2020a}, nonlinear photonics in microresonators~\cite{Moille2018,Yang2020,Zhao2020,Yu2020,Tikan2020,Lu2019,Bruch2020} and waveguides~\cite{Hickstein2018,Singh2018,Rao2019,Stanton2020}, and passive photonics and heterogeneous integration~\cite{Chang2017,Honardoost2018,Stanton2019}, and has motivated milestones such as the generation of microcombs using chip-scale lasers~\cite{Stern2018,Raja2019,Shen2020}. Photonic interposers that collect, filter, route, and interface light between many such active and passive devices are essential to realize the improvements in cost, size, weight, and power, performance, and scalability, offered by microcombs and integrated photonics, and will promote further system-level innovation using frequency combs. Such interposers need to integrate multiple broadband high-performance photonic elements, manage octave-wide light, and maintain modal and polarization purity in a low loss and high damage threshold photonics platform while pragmatically balancing heterogeneous integration and chip-to-chip coupling on a system-level architecture.

\begin{center}
\begin{figure*}
\begin{center}
\includegraphics[width=1\linewidth]{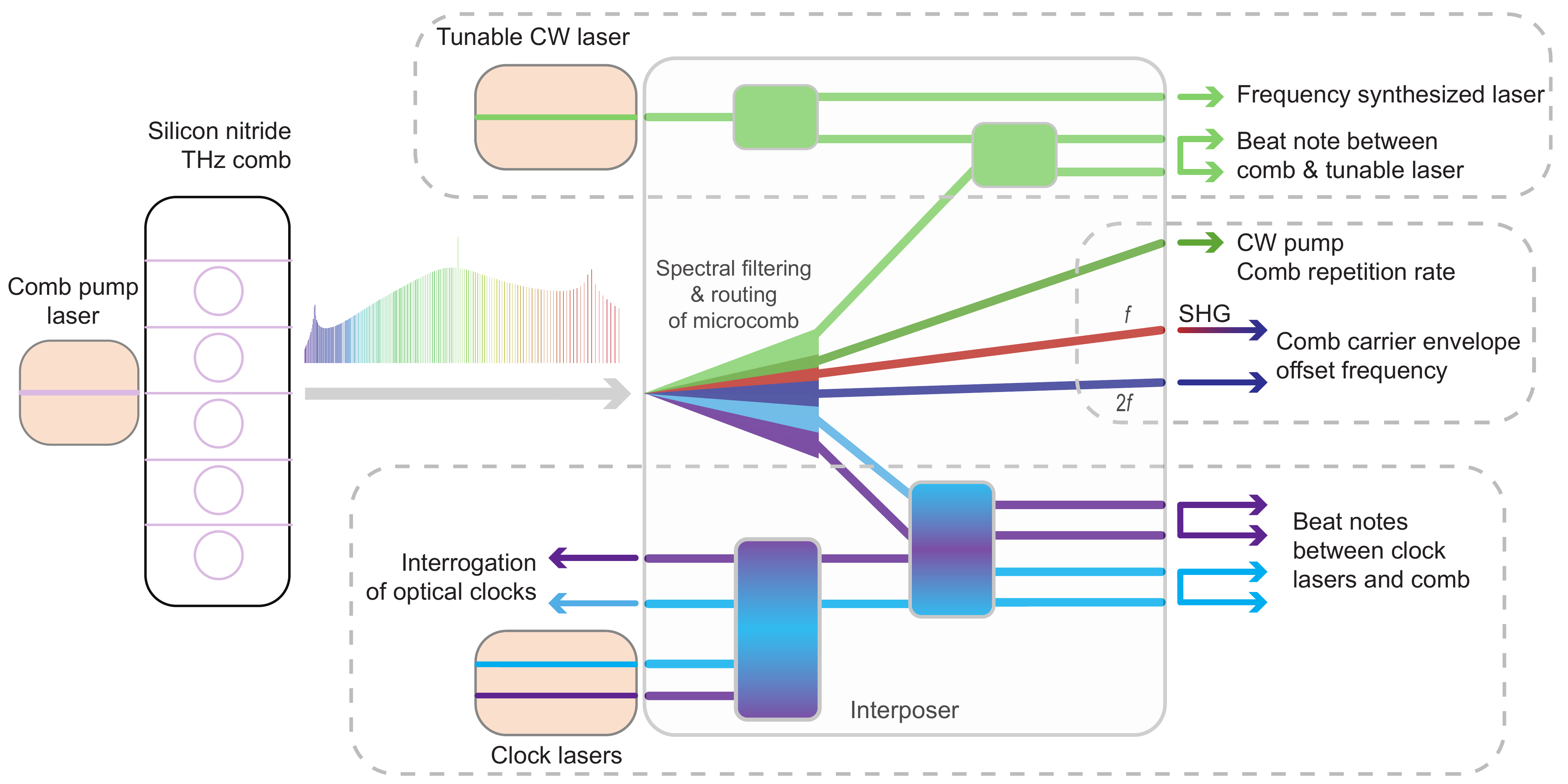}
\caption{\textbf{Photonic interposers for integrated processing of microcombs}. Photonic interposers for fully integrated microcomb-based systems need to interface multiple photonic devices, such as microcombs and other nonlinear elements, and lasers and photodetectors. The functions of the interposer can be broadly classified in two parts, first, the broadband spectral routing of microcombs, and second, the coherent mixing of specific filtered bands and teeth of the microcombs with additional external signals. The broadband spectral routing of microcombs includes separation of $f$ and 2$f$ components for self-referencing via second harmonic generation (SHG) and additional filtering for repetition rate detection, which together enable microcomb stabilization for metrological-grade applications. Depending on the application, further microcomb processing is required, such as extraction of the pertinent comb reference-band for optical frequency synthesis, or of comb teeth matched to specific atomic transitions for optical clocks. These bands and teeth are subsequently mixed with tunable lasers and clock lasers for beat note detection for synthesis and timekeeping, respectively. }
\label{fig:Fig1}
\end{center}
\end{figure*}
\end{center}

In this work, we consider integrated photonic interposers in the context of optical frequency synthesis. Optical frequency synthesis is one application in which the transition from lab-scale instrumentation to deployable technology hinges on the ability to combine microcomb technology with other integrated photonics. Optical frequency synthesizers generate stable, accurate, and precise optical frequencies from a standard microwave reference, have traditionally used mode-locked solid-state and fiber lasers to derive a fully stabilized self-referenced frequency comb~\cite{Jones2000,Holzwarth2000a}, and are indispensable in frequency metrology and timekeeping~\cite{Hall2006,Diddams2001}, coherent light detection and ranging~\cite{Swann2006}, spectroscopy~\cite{Thorpe2006}, microwave synthesis~\cite{Fortier2011}, and astronomy~\cite{Li2008}. Yet, the cost and size of such table-top systems has limited their widespread application. 

While substantial progress has been made recently towards optical frequency synthesis using integrated photonic devices~\cite{Spencer2018,Huang2016,Arafin2017a,Arafin2017,Xin2019,Singh2020}, these nascent efforts have required the use of free-space and fiber-optic components that hinders the overall goal of having standalone chip-size microcomb systems. These efforts have employed microcombs in on-chip silicon nitride and silica microresonators~\cite{Spencer2018,Huang2016} and bulk crystalline resonators~\cite{Arafin2017a}, supercontinuum and second harmonic generation in nonlinear silicon-on-insulator waveguides~\cite{Singh2020}, and phase-locking in indium phosphide photonic integrated circuits~\cite{Arafin2017a,Arafin2017}. Each of these photonic platforms offers different devices and functionality that are beneficial to building an integrated optical frequency synthesizer.

Here, we introduce an integrated photonics interposer architecture for a microcomb-based optical frequency synthesizer that collects, routes, and interfaces broadband light from discrete chiplets and heterogeneously integrated photonic devices. We use the silicon nitride (${\rm Si_3N_4}$) photonic platform, based on requirements of low absorption, high damage threshold, and broad optical transparency, and confirm that the experimental performance of  interposer elements such as dichroic couplers, resonant filters, and multimode interferometers is in agreement with our electromagnetic simulations via short-loop tests. We directly verify  the suitability of the dichroics to process octave-wide light by using an octave-spanning microcomb generated in a thick silicon nitride chip as the input. Subsequently, we demonstrate the octave-wide spectral processing of an octave-spanning microcomb, key to the interposer, via an integrated sequence of the dichroic couplers and a tunable ring filter, measuring spectral contrast between the optical bands of interest that is appropriate for our intended application and congruent with our short-loop characterization of the individual components. Further, we report the single-chip integration of a broadband ${\rm Si_3N_4}$ microcomb generated in a thick ${\rm Si_3N_4}$ layer with the thinner ${\rm Si_3N_4}$ photonic layer used for the interposer components, demonstrating a route towards additional system-level consolidation. 

\begin{center}
\begin{figure*}
\begin{center}
\includegraphics[width=1\linewidth]{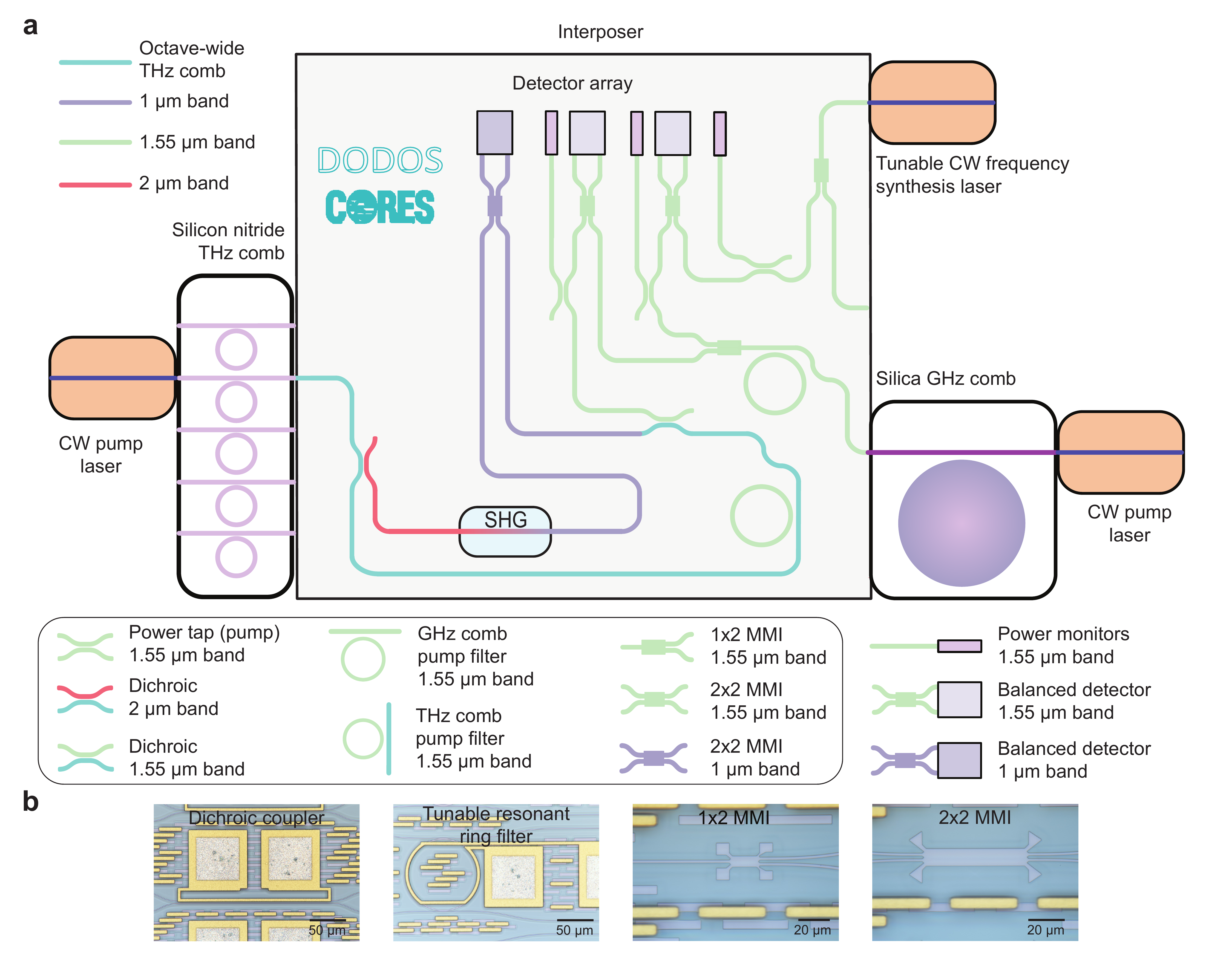}
\caption{\textbf{Interposer architecture and underlying photonic platform.} \textbf{a}. Schematic of the interposer, which is interfaced with silicon nitride and silica microcombs and a tunable laser via facet coupling, and with photodetectors and a second harmonic frequency doubler via heterogeneous integration. In the interposer, dichroic directional couplers spectrally filter the silicon nitride microcomb in preparation for self-referencing and interference with the silica microcomb for repetition rate stabilization. In turn, the output tunable frequency synthesis laser is referenced to the silica microcomb. Multimode interferometers are utilized to generate these stabilization beat notes via balanced detection, and power monitors are used for additional system-level monitoring. Here, we focus primarily on the dichroics, ring filters, and multimode interferometers. Metal traces are not shown in this schematic. \textbf{b}. Micrographs of dichroic directional couplers, ring resonator tunable filters, and multimode interferometers.}
\label{fig:Fig2}
\end{center}
\end{figure*}
\end{center}

Figure~1 schematically depicts microcombs and other integrated photonic devices in the context of systems such as optical frequency synthesizers and optical atomic clocks. To transition to an integrated system from the table-top, numerous optical functions are required with the simultaneous operation of multiple photonic devices in lockstep. These functions nominally translate to different materials requirements - optical gain is required for lasers, ${\rm \chi^{(3)}}$ nonlinearity for microcombs, ${\rm \chi^{(2)}}$ nonlinearity for second harmonic generation, low linear loss for passives, and a high responsivity, low dark current material for photodetectors. To address this challenge of combining multiple material responses and platforms, one approach is to interface several chiplets of different photonic materials on a common carrier via chip-to-chip facet coupling, benefiting from the use of reliable well-established photonics and the ability to prequalify each photonic element prior to system assembly. Another approach is to integrate all functions and materials together on one main photonic chip, akin to heterogeneous integration, where the benefits inherent to having a system on a chip will come at the cost of the requisite research and development. Crucially, a judicious combination of chip-to-chip facet coupling and heterogeneous integration can balance the pragmatism of using discrete chiplets of well-established photonic elements with the benefits and cost of heterogeneously integrating multiple material systems together, using a photonic interposer to bind the system together.

\begin{center}
\begin{figure*}
\begin{center}
\includegraphics[width=1\linewidth]{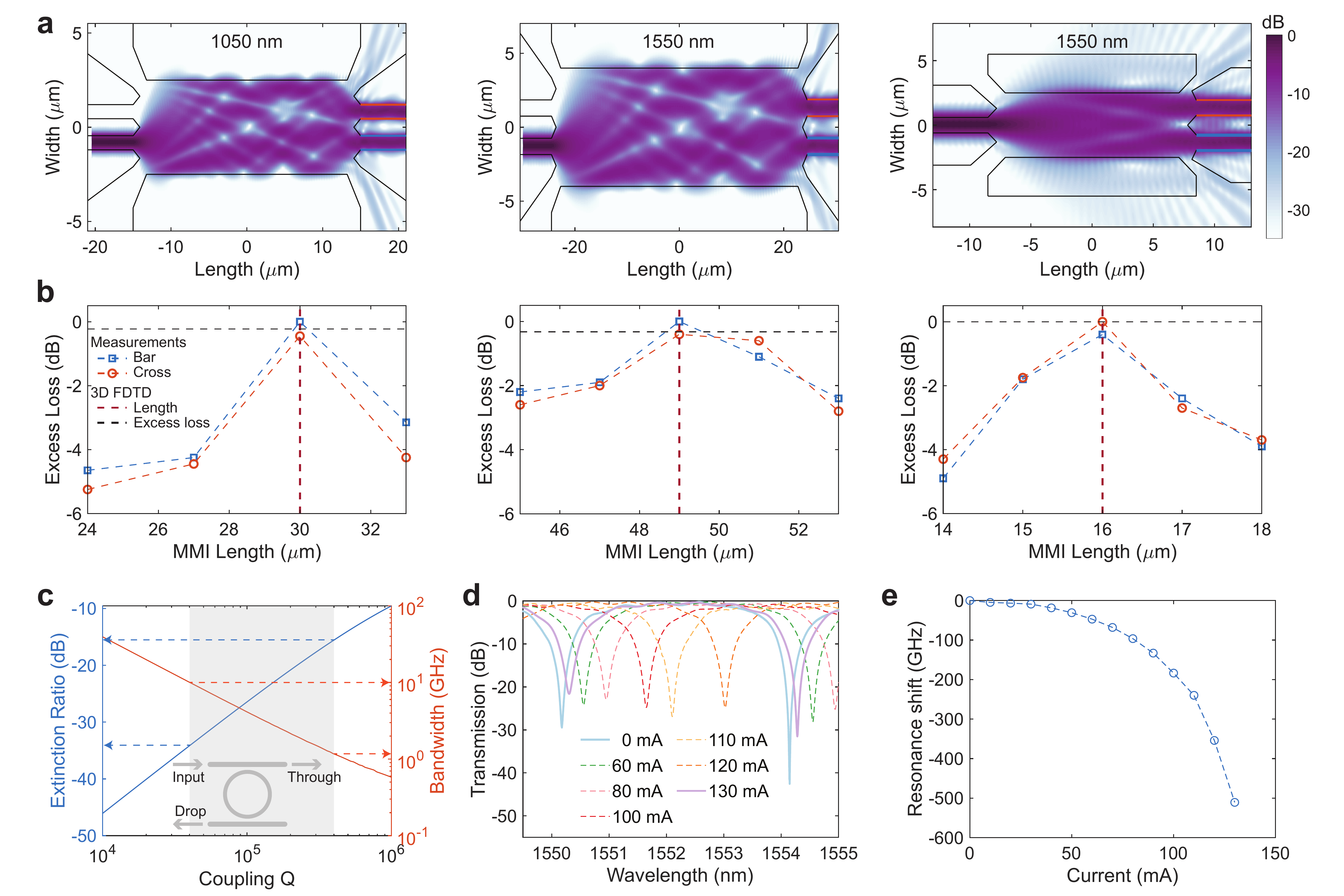}
\caption{\textbf{Multimode interferometers and microring filters}. \textbf{a}. Simulations showing the propagation of light from left to right in the three multimode interferometers at 1050~nm and 1550~nm. The corners of the butterfly geometry guide out light at the $\approx$ -25~dB level, suppressing potential reflections. The bar and cross output ports are highlighted in orange and blue outlines, respectively. Cross-sections of $ \lvert \textbf{E}(x,y,z)\rvert^2$ are plotted with z set to half the height of the MMIs. \textbf{b}. Corresponding continuous-wave measurements of the bar and cross ports of the MMIs for a range of MMI lengths. In each case, the optimal MMI length matches the predicted length from the simulations in \textbf{a}. The associated measurement uncertainty is less than 0.2~dB based on one standard deviation in the transmission of five identical cascaded multimode interferometers. \textbf{c}. Simulated dependence of microring filter characteristics, extinction ratio and bandwidth, on coupling~Q for an intrinsic~Q  of ${\rm 10^6}$. Coupling~Qs between 4$\times10^4$ and 4$\times10^5$ yield extinction ratios between 15~to~35~dB and corresponding filter bandwidths of 1~to~10~GHz, a range of filter characteristics suitable for our intended application of suppressing the pump of microcombs. \textbf{d}. Measured transmission spectra for a thermally tuned microring filter using an integrated heater. \textbf{e}. Variation of resonance frequency shift with heater current corresponding to \textbf{d}, showing over one free spectral range of tuning.}
\label{fig:Fig3}
\end{center}
\end{figure*}
\end{center}

Figure~1 also indicates the nature of microcomb processing required of such photonic interposers. Spectral bands of combs generated in nonlinear resonators pumped by chip-scale lasers need to be adequately filtered across an octave bandwidth to facilitate stabilization via f-2f self-referencing, where additional nonlinear devices are required for the frequency doubling. Additionally, narrow spectral filtering of the strong pumps that drive the microcombs is required to prevent damage to and maintain the performance of both slow and fast photodetectors that monitor optical power and facilitate phase-locking via optical interference in on-chip coherent mixers. The approach upon which our interposer design is based uses two phase-stable interlocking Kerr combs to form the optical reference for synthesis~\cite{Spencer2018}, each pumped near 1550~nm, and generated in separate silicon nitride (${\rm Si_3N_4}$) and silica (${\rm SiO_2}$) microresonators with repetition rates of $\approx$ 1 THz and $\approx$ 20~GHz, respectively. The dual-microcomb system assists in reducing power consumption and facilitates ease of operation, where the octave spanning ${\rm Si_3N_4}$ comb is used for self-referencing and a narrower ${\rm SiO_2}$ comb is used for repetition rate and synthesis frequency detection.

Figure~2a shows a schematic of the full photonic interposer design, which is based on transverse-electric (TE) polarized guided light in a 400~nm thick stoichiometric ${\rm Si_3N_4}$ photonic platform with upper and lower silicon dioxide cladding. The ${\rm Si_3N_4}$ platform is well established for numerous applications, and its low optical loss and high optical damage threshold, coupled with its broad optical transparency, assist in processing both low and high-power optical signals across the octave bandwidth. The nitride film thickness and waveguide widths are chosen to balance optical confinement, proximity to the optical single mode condition, and coupling to both heterogeneously integrated and facet-coupled elements, in contrast to microcombs where the anomalous dispersion required for octave spanning bright Kerr solitons necessitates films that are nearly a factor of two thicker. Further details regarding optical confinement and the number of modes can be found in the Supplementary Information.

The interposer is comprised of dichroic directional couplers (hereafter referred to as dichroics), resonant filters, 50:50 multimode interferometers (MMIs), and power splitters and taps that operate on the two microcombs and the tunable synthesis laser (Fig.~2b). These elements interface with a frequency doubler (SHG), polarization rotator, and photodetector array that are heterogeneously integrated and whose design and performance have been demonstrated elsewhere~\cite{Stanton2019,Stanton2020,Yu2020a,Costanzo2020}. The output of the octave-spanning ${\rm Si_3N_4}$ comb chip is directed to two cascaded dichroics that spectrally filter the microcomb into three key spectral bands, a long and a short wavelength band around 2~${\rm \mu}$m and 1~${\rm \mu}$m respectively, separated by an octave, and the center band around 1.55~${\rm \mu}$m. The first dichroic separates out light in the 2 ${\rm \mu}$m band from shorter wavelengths, and the second dichroic separates 1.55~${\rm \mu}$m light from shorter wavelengths (in particular, the 1~${\rm \mu}$m light). The 2~${\rm \mu}$m light is led to the frequency doubler, after which the upconverted output in the 1~${\rm \mu}$m band is coherently mixed with the 1~${\rm \mu}$m microcomb light in a 2$\times$2~50:50~MMI and detected to extract the carrier envelope offset frequency of the THz comb. Two 1$\times$2~50:50~MMIs split the ${\rm SiO_2}$ comb and the tunable synthesis laser (which reside on separate chips that are butt-coupled to the interposer). An additional 2$\times$2~50:50~MMI is used to coherently mix the ${\rm SiO_2}$ comb with the 1.55 ${\rm \mu}$m band of the ${\rm Si_3N_4}$ comb light, while a second 2$\times$2 50:50 MMI mixes the ${\rm SiO_2}$ comb with the tunable laser. The MMI outputs are used to phase-lock the two microcombs and detect the precise optical frequency of the tunable laser. In addition, two thermally tunable microring resonators filter out the microcomb pumps in the 1.55~${\rm \mu}$m band, and power taps and detectors are used to monitor the optical power of the microcombs and tunable laser. We design these interposer components employing a combination of waveguide eigenmode and 3D finite-difference time-domain (FDTD) simulations, and fabricate them on 100~mm wafers using process sequences based on both deep ultraviolet lithography (Ligentec) and electron-beam lithography (NIST). We validate our designs and fabrication by experimentally confirming the predicted component performance using both continuous-wave (CW) light and octave spanning microcomb light. 

Figure~3a shows 3D FDTD simulations of the 1$\times$2 and 2$\times$2 50:50 MMIs that function as power splitters and coherent mixers, respectively (see Supplementary Information for details). The transmission ratio of the optical powers at the output ports of the 2$\times$2 MMIs impact the balanced detection of the beat notes for phase-locking, motivating our choice of a butterfly multimode interferometer over a directional coupler. The corners of the butterfly geometry funnel out potential reflections that are deleterious to both the unity transmission ratio and the operation of an integrated circuit~\cite{Kleijn2014}. The corresponding CW transmission measurements of the bar and cross ports are shown in Fig.~3b for a range of MMI lengths, and the optimum MMI length agrees with our simulations. The excess loss, defined as transmission loss relative to the maximum transmission (nominally -3~dB), for all three optimal MMIs lengths is less than 0.5~dB, and includes variations from coupling on and off the chip.

The filter bandwidth and extinction ratio of the thermally tunable symmetric add-drop microring filters that filter CW pump light are determined by the intrinsic and coupling quality factors (Q), which depend on absorption and scattering, and on the magnitude of coupling between the bus waveguide and the microring~\cite{Manolatou1999,Moille2019}, respectively (Fig.~3c). Measurements (Fig.~3d and 3e) show that a ring filter with 50~$\mu$m radius (474.8 GHz free spectral range) suitable for the ${\rm Si_3N_4}$ microcomb (coupling~Q $\approx$ 2$\times10^4$, intrinsic Q $\approx10^{6}$) can be thermally tuned over 500 GHz, i.e., over an entire free spectral range, while maintaining adequate extinction, a requirement for matching the resonance of the filter with the pump of the ${\rm Si_3N_4}$ microcomb. The maximum extinction measured, and variations therein, are limited by thermally induced perturbations to the coupling, and the polarization extinction ratio of the input light. For typical THz repetition rate microcombs, the pump power is 15 to 20 dB higher than the neighboring comb teeth. Therefore, to flatten the pump comb tooth to match the surrounding teeth, a coupling Q as high as  $\approx 10^5$ can be adequate. A similar microring with coupling Q $\approx 10^5$ will be suitable for filtering the ${\rm SiO_2}$ microcomb. Additional details regarding design and fabrication can be found in the Supplementary Information. While our intended application requires moderate filtering and can take advantage of an inherent vernier effect between the filter and microcomb resonators, more demanding applications can use cascaded ring filters to synthesize more complex filter responses~\cite{Barwicz2004,Amatya2008}.

\begin{center}
\begin{figure*}
\begin{center}
\includegraphics[width=1\linewidth]{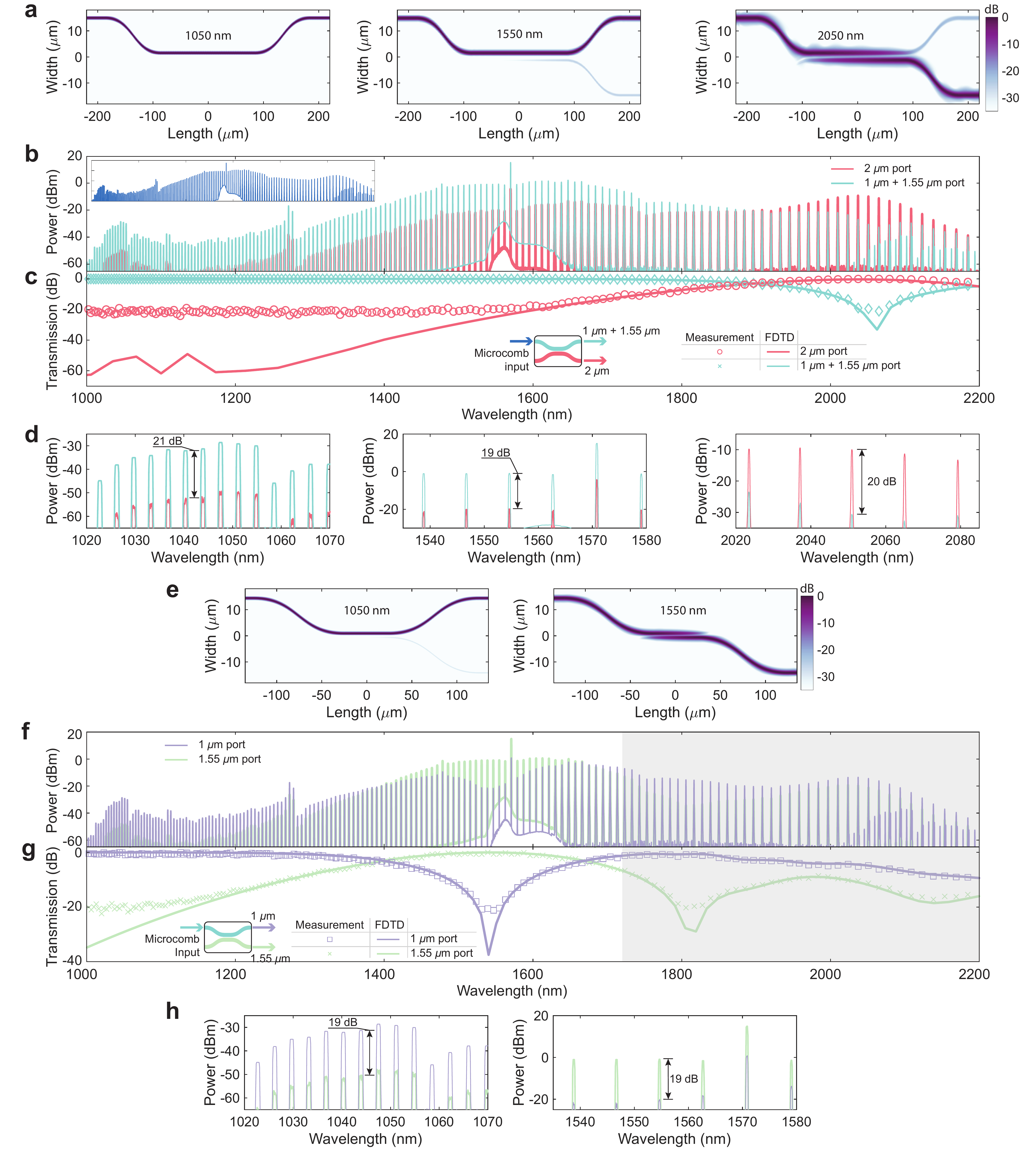}
\caption{\textbf{Octave wide operation of dichroics}. \textbf{a}-\textbf{d} First dichroic, whose purpose is to separate 2~$\mu$m light from shorter wavelengths. \textbf{a}. Simulations at 1050 nm, 1550 nm, and 2050 nm showing extraction of the 2 ${\rm \mu}$m band into the cross port. \textbf{b}. Measured broadband experimental spectra at the bar and cross ports. The input is the microcomb shown in the inset. \textbf{c}. Measured (symbols) and simulated (solid lines) octave wide transfer function. At the cross or 2~$\mu$m~port, extinction ratios of (${\rm 21.4~\pm~1.1}$)~dB and (${\rm 19.9~\pm~0.8}$)~dB are measured in the 1~${\rm \mu}$m and 1.55~${\rm \mu}$m bands, respectively. At the bar port, an extinction ratio of (${\rm 18.1~\pm~2.9}$)~dB is measured in the 2~${\rm \mu}$m band. \textbf{d}. Magnified individual spectral bands. \textbf{e}-\textbf{h} Second dichroic, whose purpose is to separate 1.55~$\mu$m light from shorter wavelengths. \textbf{e}. Simulations at 1050 nm and 1550 nm, showing extraction of the 1.55~${\rm \mu}$m band into the cross port.  
\textbf{f}. Measured broadband experimental spectra at the bar and cross ports. The input is the microcomb shown in the inset of \textbf{b}.
\textbf{g}. Measured (symbols) and simulated (solid lines) octave wide transfer function. At the cross or 1.55~$\mu$m~port, an extinction ratio of (${\rm 20.1~\pm~1.0}$)~dB is measured in the 1~${\rm \mu}$m band, and at the bar port, an extinction ratio of (${\rm 18.6~\pm~3.3}$)~dB is measured in the 1.55~${\rm \mu}$m band. \textbf{d}. Magnified individual spectral bands. The performance of the dichroic in the spectral region shaded in \textbf{f} and \textbf{g} is relatively unimportant, as this region is filtered out by the first dichroic in the full interposer chip. In  \textbf{a} and \textbf{e}, cross-sections of $ \lvert \textbf{E}(x,y,z)\rvert^2$ are plotted with z set to half the height of the dichroics. The measured transfer functions shown in \textbf{c} and \textbf{g} are extracted from the corresponding transmission of the comb teeth in \textbf{b} and \textbf{f}. The corresponding uncertainties reported in \textbf{c} and \textbf{g} correspond to line-to-line fluctuations in the measured comb spectra and include variations in coupling and are one standard deviation values.}
\label{fig:Fig4}
\end{center}
\end{figure*}
\end{center}

\begin{center}
\begin{figure*}
\begin{center}
\includegraphics[width=1\linewidth]{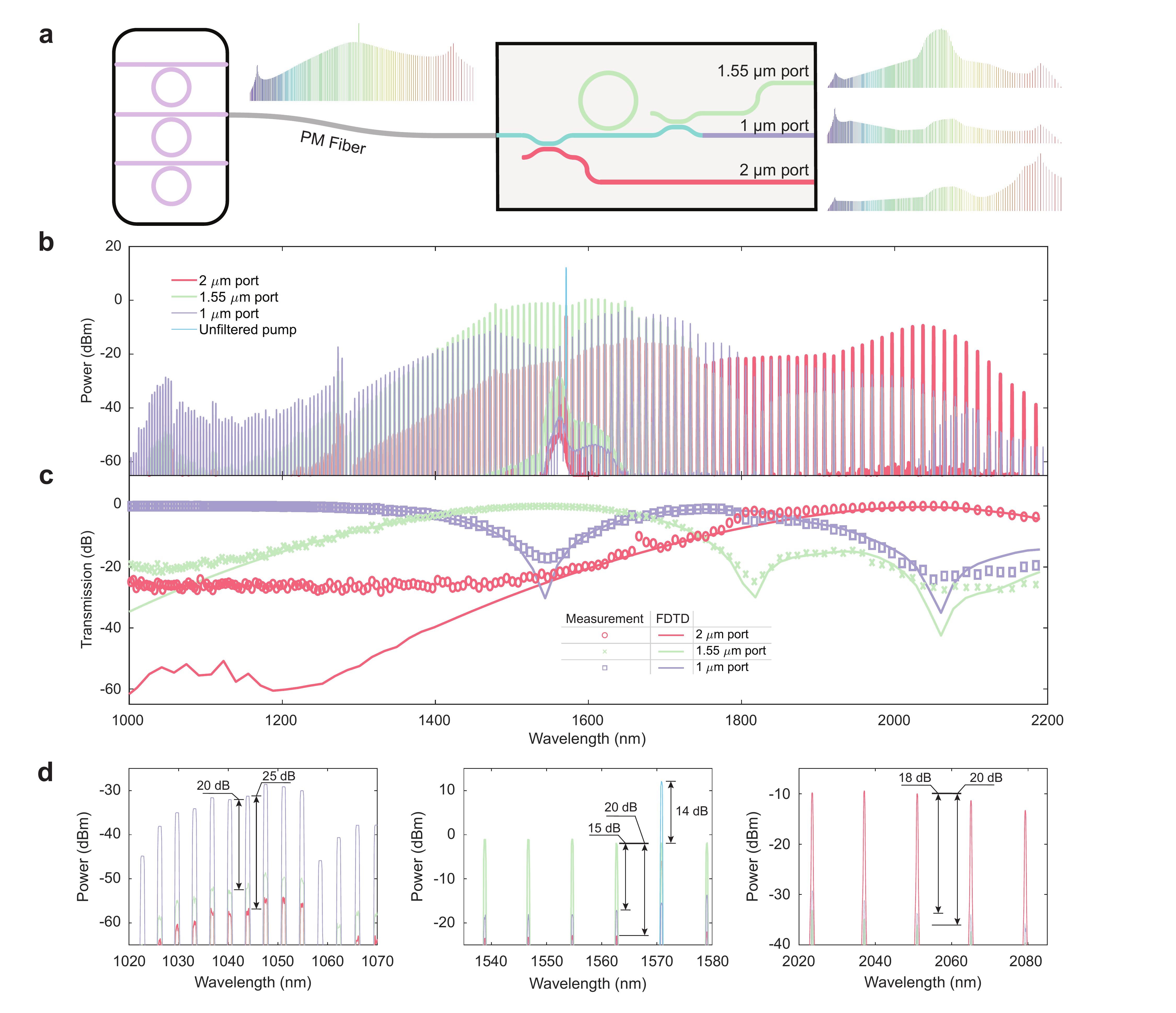}
\caption{\textbf{Integrated spectral processing of a microcomb}. \textbf{a}. Schematic for on-chip processing of a silicon nitride based octave spanning microcomb. PM = polarization maintaining. \textbf{b}. Experimental spectra measured at the three output ports. The microcomb shown in the inset of Fig.~4\textbf{b} is used as the input. \textbf{c}. Measured (symbols) and simulated (solid lines) octave wide transfer functions. The measured transfer function is extracted from the transmission of the comb teeth in \textbf{b}. At the 1~$\mu$m~port, extinction ratios of (${\rm 16.2~\pm~0.8}$)~dB and (${\rm 20.9~\pm~2.2}$)~dB are measured in the 1.55~${\rm \mu}$m and 2~${\rm \mu}$m bands, respectively. Similarly, at the 1.55~$\mu$m~port, extinction ratios of (${\rm 20.2~\pm~0.7}$)~dB and (${\rm 25.6~\pm~2.1}$)~dB are measured in the 1~${\rm \mu}$m and 2~${\rm \mu}$m bands, and at the 2~$\mu$m~port, extinction ratios of (${\rm 26.1~\pm~0.8}$)~dB and (${\rm 22.7~\pm~0.9}$)~dB are measured in the 1~${\rm \mu}$m and 1.55~${\rm \mu}$m bands. \textbf{d}. Magnified comparison of the outputs at the three ports in the individual spectral bands. Separation of the three spectral bands into the three ports with 15~dB to 25~dB of contrast is observable, along with 14~dB of pump suppression after comb generation from the ring filter (light blue comb tooth). The uncertainties reported in \textbf{c} correspond to line-to-line fluctuations in the comb spectra and include variations in coupling, and are one standard deviation values.}
\label{fig:Fig5}
\end{center}
\end{figure*}
\end{center}

Figure~4a shows simulations for the dichroic that extracts the 2~${\rm \mu}$m microcomb band into the cross port. We measured the cross and bar port transmission for a range of directional coupler lengths using CW light at the three bands, and observed agreement with the expected optimized coupler length, with 15~dB of contrast at 2~${\rm \mu}$m and 1.55~${\rm \mu}$m, and over 30~dB at 1~${\rm \mu}$m, see the Supplementary Information for details. Figure~4b shows the measured individual bar and cross port spectra of the optimized dichroic across the nominal octave bandwidth centered around the telecom C band. The measurement uses an octave spanning ${\rm Si_3N_4}$ microcomb (Fig.~4b, inset), generated in a 770~nm thick microring with low and broadband anomalous dispersion, as the input. Figure~4c compares the measured transmission with the simulated transmission, and magnified views of measurements in the 2~${\rm \mu}$m, 1.55~${\rm \mu}$m, and 1~${\rm \mu}$m bands are shown in Fig.~4d. Similarly, the second dichroic couples out the 1.55~${\rm \mu}$m microcomb light into the cross port, leaving the 1~${\rm \mu}$m band in the bar port, as seen in simulations at these wavelengths in Fig.~4e. Corresponding CW measurements indicated over 20~dB of contrast between the two ports, see the Supplementary Information for details. The behavior of this dichroic in the 2 ${\rm \mu}$m band is inconsequential because it is intended to process the ${\rm Si_3N_4}$ microcomb after the 2~${\rm \mu}$m band is filtered out in the first dichroic (Fig.~2). Figure~4f shows the measured individual bar and cross port spectra of the optimized dichroic, using the same microcomb input employed to evaluate the first dichroic (Fig~4b, inset). Figure~4g compares the simulated and measured transmission of the dichroic across the octave, and magnified views of the spectral bands are shown in Fig.~4h. Overall, the performance of the two dichroics is appropriate for our intended application and largely follows the simulated behavior, with deviations observed only below the $\approx$ -20~dB level, likely originating from limitations of the measurement setup. Further details regarding design optimization and the experimental setup can be found in the Supplementary Information.

\begin{center}
\begin{figure*}
\begin{center}
\includegraphics[width=1\linewidth]{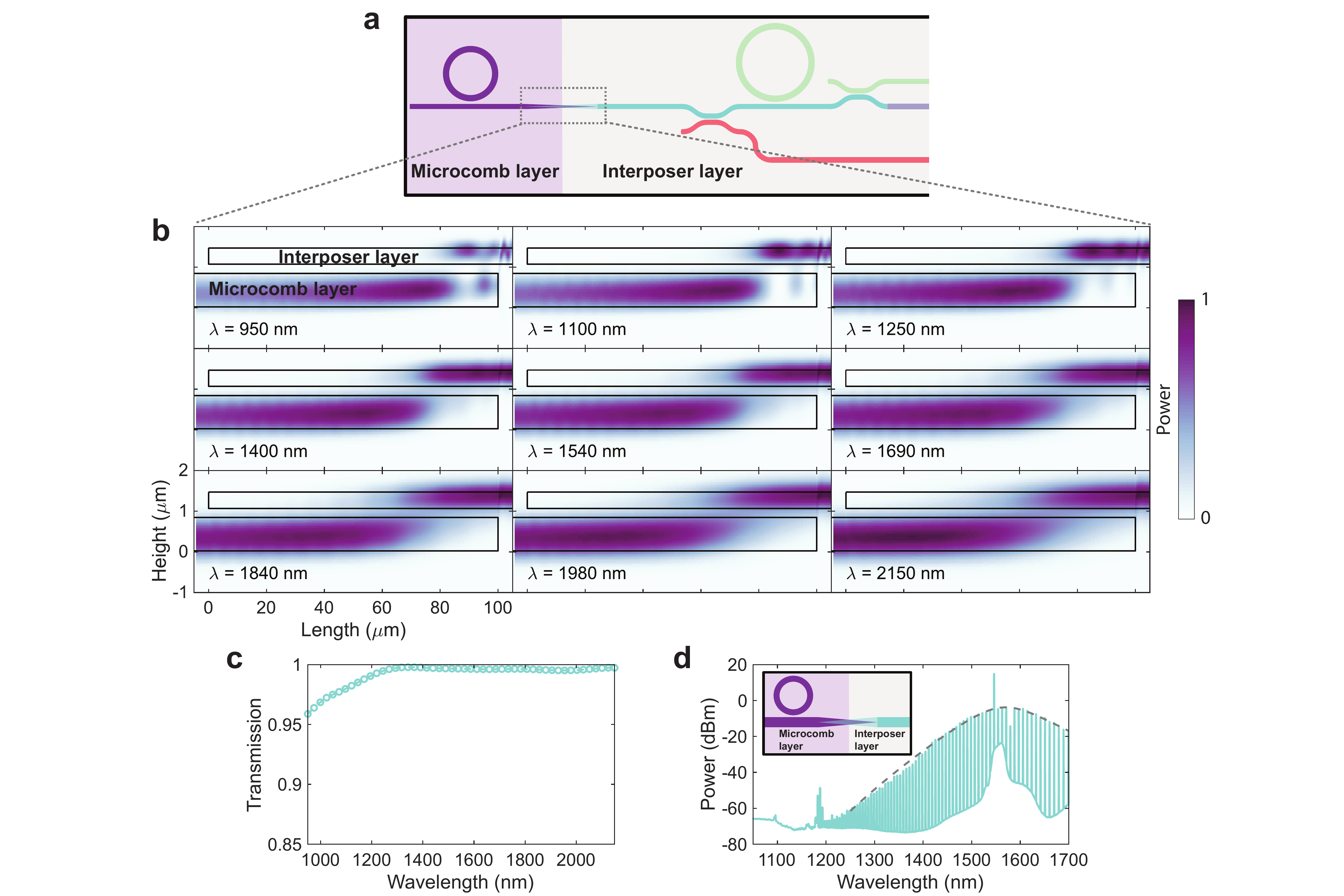}
\caption{\textbf{Towards microcomb-interposer integration}. \textbf{a}. Vision for microcomb-interposer integration, where the two photonic layers are interfaced by a bilayer taper that can transfer an octave of comb bandwidth with negligible loss. \textbf{b},\textbf{c}. Simulations of a 100~${\rm \mu}$m long bilayer taper showing low-loss broadband transfer of light. \textbf{d}. Broadband microcomb, along with the corresponding sech$^2$ fit, measured from a fabricated bilayer chip where the microcomb output is extracted through the bilayer taper into the interposer layer.}
\label{fig:Fig6}
\end{center}
\end{figure*}
\end{center}

So far, we have presented the design and experimental characterization of individual interposer elements. As a first demonstration of processing an octave spanning microcomb using a more integrated photonic chip that contains all of the aforementioned filtering capability, we measured the transmission through a chip comprised of a sequence of the two dichroics with a microring filter at the 1.55~${\rm \mu}$m band port (Fig.~5a), using an octave spanning microcomb (Fig.~4b, inset) as the input. Measurements shown in Fig.~5b show that the three spectral bands of interest are routed into the three physical ports. The ring filter reduces the pump amplitude to that of the neighboring comb tones. Figure~5c compares the transfer function extracted from Fig.~5b to simulations based on 3D~FDTD, excluding the effect of the ring filter that has no effect on the transmission envelope, showing good agreement between the two. Magnified views of the three spectral bands are shown in Fig.~5d. We observe 15~dB to 25~dB of extinction across the spectral bands at the outputs, along with 14~dB of pump suppression from the ring filter. Similar to the characterization of the individual dichroics, deviations occurring below the $\approx$ -20~dB level result from limitations of the measurement; see the Supplementary Information for more details regarding the fabrication and experimental setup.

Looking forward, we show that our microcomb sources can be integrated with our photonic interposer layer, as envisioned in Fig.~6a. Bright Kerr soliton generation directly within the 400~nm thick Si$_3$N$_4$ interposer layer is not possible in conventional ring geometries due to the normal dispersion associated with all waveguide widths at that thickness. One could instead consider making the interposer out of a thicker Si$_3$N$_4$ layer (i.e., suitable for broadband anomalous dispersion), but the design of passive elements may be complicated by the increased confinement and larger numbers of modes supported by the thicker film. We instead adopt a dual layer approach, shown in Fig.~6. Here, fabrication of a thick Si$_3$N$_4$ layer (the microcomb layer) is followed by a vertically-coupled thin Si$_3$N$_4$ layer (the interposer layer), with chemical-mechanical polishing enabling control of the SiO$_2$ film thickness separating the layers. A key challenge for this approach is the transfer of the microcomb to the interposer layer across a full octave of bandwidth. We address this challenge by using a 100~${\rm \mu}$m bilayer taper (schematic top-view shown in Fig.~6d) that ensures adiabatic transfer of light with less than 1~dB of loss across an octave, simulated using 3D~FDTD (Fig.~6b and 6c). The Si$_3$N$_4$ film thicknesses of the microcomb and interposer layers are 790~nm (a common thickness for broadband combs~\cite{Marin-Palomo2017,Riemensberger2020}) and 400~nm, respectively, with an interlayer SiO$_2$ thickness of 200~nm (see Supplementary Information for details). Both layers are tapered in width from 1~${\rm \mu}$m to 0.2~${\rm \mu}$m over a 100~${\rm \mu}$m length. Importantly, the adiabatic nature of the taper is such that it is relatively insensitive to precise interlayer SiO$_2$ thickness (at the 50~nm level), as well as lateral offsets between the waveguide layers (at the 100~nm level). Figure~6d shows a Kerr soliton microcomb generated in a ring of 23~${\rm \mu}$m radius, measured after transfer through the bilayer taper. No spectral degradation was observed in comparison to a microcomb pumped in the opposite direction, where the microcomb does not pass through the bilayer taper. The reduced bandwidth of the microcomb compared to that used previously in this work stems from differences in dispersion that arise from the different Si$_3$N$_4$ thickness used (790~nm here vs. 770~nm previously). Nevertheless, this serves as a conclusive demonstration that the thick Si$_3$N$_4$ layer associated with microcomb generation can be integrated on the same chip with thinner Si$_3$N$_4$ that is more optimal for linear functionality.

In summary, we have introduced an integrated photonics interposer architecture for a microcomb-based optical frequency synthesizer that uses a variety of photonic devices to collect, route, and interface broadband light from discrete chiplets and heterogeneously integrated photonic devices, addressing a key impediment in the full chip-scale integration of multiple material systems and functional responses for microcomb-based systems. We use the well-known ${\rm Si_3N_4}$ photonic platform because of its low absorption, high damage threshold, and broad optical transparency, and validate our approach with a combination of electromagnetic calculations and measurements on fabricated devices integral to the interposer. We perform a series of short-loop tests where our designs for dichroic couplers, resonant filters, and multimode interferometers show experimental performance well-suited for processing microcombs, in congruence with our simulations. In addition to measurements using continuous-wave inputs, we use an octave-spanning microcomb generated in a thick silicon nitride chip as the input to directly confirm the ability of the dichroic couplers to process octave-wide light. Following the success of the interposer elements, we implement the first demonstration of octave-wide spectral processing of an octave-spanning microcomb through an integrated chain of two dichroic couplers and a ring filter which constitute the key broadband comb processing sequence of the interposer, and measure 15~dB to 25~dB of spectral contrast in the wavelength bands of interest, along with the flattening of the pump tone to match the remainder of the microcomb. Further, we report the single-chip integration of a broadband ${\rm Si_3N_4}$ microcomb with the ${\rm Si_3N_4}$ photonic layer used for the interposer components by using a broadband adiabatic taper to transfer the microcomb output between the thick microcomb and thinner interposer layers, indicating a path towards integrating microcombs with additional customizable photonic processing. The interposer architecture introduced here can be adapted to other microcomb-based integrated systems for optical atomic clocks, high-precision spectroscopy, and precise navigation, among others, based on similar requirements for microcomb-processing and system integration. 
\\

%\bibliographystyle{osajnl}
%\bibliography{Interposer_references.bib}

\begin{thebibliography}{10}
\newcommand{\enquote}[1]{``#1''}

\bibitem{Cundiff2003}
S.~T. Cundiff and J.~Ye, \enquote{{Colloquium: Femtosecond optical frequency
  combs},} {{Reviews of Modern Physics}} \textbf{75},
  325--342 (2003).

\bibitem{Kippenberg2018a}
T.~J. Kippenberg, A.~L. Gaeta, M.~Lipson, and M.~L. Gorodetsky,
  \enquote{{Dissipative Kerr solitons in optical microresonators},}
  { {Science}} \textbf{361}, eaan8083 (2018).

\bibitem{Gaeta2019}
A.~L. Gaeta, M.~Lipson, and T.~J. Kippenberg, \enquote{{Photonic-chip-based
  frequency combs},} { {Nature Photonics}} \textbf{13},
  158--169 (2019).

\bibitem{Pasquazi}
A.~Pasquazi, M.~Peccianti, L.~Razzari, D.~J. Moss, S.~Coen, M.~Erkintalo, Y.~K.
  Chembo, T.~Hansson, S.~Wabnitz, P.~Del'Haye, X.~Xue, A.~M. Weiner, and
  R.~Morandotti, \enquote{{Micro-combs: A novel generation of optical
  sources},} { {Physics Reports}} \textbf{729}, 1--81
  (2018).

\bibitem{Moille2018}
G.~Moille, Q.~Li, S.~Kim, D.~Westly, and K.~Srinivasan, \enquote{{
  Phased-locked two-color single soliton microcombs in dispersion-engineered
  Si$_{3}$N$_{4}$ resonators },} { {Optics Letters}}
  \textbf{43}, 2772--2775 (2018).

\bibitem{Yang2020}
J.~Yang, S.~W. Huang, Z.~Xie, M.~Yu, D.~L. Kwong, and C.~W. Wong,
  \enquote{{Coherent satellites in multispectral regenerative frequency
  microcombs},} { {Communications Physics}} \textbf{3},
  1--9 (2020).

\bibitem{Zhao2020}
Y.~Zhao, X.~Ji, B.~Y. Kim, P.~S. Donvalkar, J.~K. Jang, C.~Joshi, M.~Yu,
  C.~Joshi, R.~R. Domeneguetti, F.~A.~S. Barbosa, P.~Nussenzveig, Y.~Okawachi,
  M.~Lipson, and A.~L. Gaeta, \enquote{{Visible nonlinear photonics via
  high-order-mode dispersion engineering},} { {Optica}}
  \textbf{7}, 135--141 (2020).

\bibitem{Yu2020}
S.-P. Yu, D.~C. Cole, H.~Jung, G.~T. Moille, K.~Srinivasan, and S.~B. Papp,
  \enquote{{Spontaneous Pulse Formation in Edge-Less Photonic Crystal
  Resonators},} { {arXiv:2002.12502}}  (2020).

\bibitem{Tikan2020}
A.~Tikan, J.~Riemensberger, K.~Komagata, S.~H{\"{o}}nl, M.~Churaev, C.~Skehan,
  H.~Guo, R.~N. Wang, J.~Liu, P.~Seidler, and T.~J. Kippenberg,
  \enquote{{Emergent Nonlinear Phenomena in a Driven Dissipative Photonic
  Dimer},} { {arXiv:2005.06470}}  (2020).

\bibitem{Jones2000}
D.~J. Jones, S.~A. Diddams, J.~K. Ranka, A.~Stentz, R.~S. Windeler, J.~L. Hall,
  and S.~T. Cundiff, \enquote{{Carrier-envelope phase control of femtosecond
  mode-locked lasers and direct optical frequency synthesis},}
  { {Science}} \textbf{288}, 635--639 (2000).

\bibitem{Holzwarth2000a}
R.~Holzwarth, T.~Udem, T.~W. H{\"{a}}nsch, J.~C. Knight, W.~J. Wadsworth, and
  P.~S.~J. Russell, \enquote{{Optical frequency synthesizer for precision
  spectroscopy},} { {Physical Review Letters}} \textbf{85},
  2264--2267 (2000).

\bibitem{Spencer2018}
D.~T. Spencer, T.~Drake, T.~C. Briles, J.~Stone, L.~C. Sinclair, C.~Fredrick,
  Q.~Li, D.~Westly, B.~R. Ilic, A.~Bluestone, N.~Volet, T.~Komljenovic,
  L.~Chang, S.~H. Lee, D.~Y. Oh, M.~G. Suh, K.~Y. Yang, M.~H. Pfeiffer, T.~J.
  Kippenberg, E.~Norberg, L.~Theogarajan, K.~Vahala, N.~R. Newbury,
  K.~Srinivasan, J.~E. Bowers, S.~A. Diddams, and S.~B. Papp, \enquote{{An
  optical-frequency synthesizer using integrated photonics},}
  { {Nature}} \textbf{557}, 81--85 (2018).

\bibitem{Diddams2001}
S.~A. Diddams, T.~Udem, J.~C. Bergquist, E.~A. Curtis, R.~E. Drullinger,
  L.~Hollberg, W.~M. Itano, W.~D. Lee, C.~W. Oates, K.~R. Vogel, and D.~J.
  Wineland, \enquote{{An optical clock based on a single trapped 199Hg+ ion},}
  { {Science}} \textbf{293}, 825--828 (2001).

\bibitem{Newman2019}
Z.~L. Newman, V.~Maurice, T.~Drake, J.~R. Stone, T.~C. Briles, D.~T. Spencer,
  C.~Fredrick, Q.~Li, D.~Westly, B.~R. Ilic, B.~Shen, M.-G. Suh, K.~Y. Yang,
  C.~Johnson, D.~M.~S. Johnson, L.~Hollberg, K.~J. Vahala, K.~Srinivasan, S.~A.
  Diddams, J.~Kitching, S.~B. Papp, and M.~T. Hummon, \enquote{{Architecture
  for the photonic integration of an optical atomic clock},}
  { {Optica}} \textbf{6}, 680--685 (2019).

\bibitem{Swann2006}
W.~C. Swann and N.~R. Newbury, \enquote{{Frequency-resolved coherent lidar
  using a femtosecond fiber laser},} { {Optics Letters}}
  \textbf{31}, 826--828 (2006).

\bibitem{Suh2018}
M.~G. Suh and K.~J. Vahala, \enquote{{Soliton microcomb range measurement},}
  { {Science}} \textbf{359}, 884--887 (2018).

\bibitem{Riemensberger2020}
J.~Riemensberger, A.~Lukashchuk, M.~Karpov, W.~Weng, E.~Lucas, J.~Liu, and
  T.~J. Kippenberg, \enquote{{Massively parallel coherent laser ranging using a
  soliton microcomb},} { {Nature}} \textbf{581}, 164--170
  (2020).

\bibitem{Thorpe2006}
M.~J. Thorpe, K.~D. Moll, J.~R. Jones, B.~Safdi, and J.~Ye, \enquote{{Broadband
  cavity ringdown spectroscopy for sensitive and rapid molecular defection},}
  { {Science}} \textbf{311}, 1595--1599 (2006).

\bibitem{Suh2016}
M.~G. Suh, Q.~F. Yang, K.~Y. Yang, X.~Yi, and K.~J. Vahala,
  \enquote{{Microresonator soliton dual-comb spectroscopy},}
  { {Science}} \textbf{354}, 600--603 (2016).

\bibitem{Dutt2018}
A.~Dutt, C.~Joshi, X.~Ji, J.~Cardenas, Y.~Okawachi, K.~Luke, A.~L. Gaeta, and
  M.~Lipson, \enquote{{On-chip dual-comb source for spectroscopy},}
  { {Science Advances}} \textbf{4}, e1701858 (2018).

\bibitem{Fortier2011}
T.~M. Fortier, M.~S. Kirchner, F.~Quinlan, J.~Taylor, J.~C. Bergquist,
  T.~Rosenband, N.~Lemke, A.~Ludlow, Y.~Jiang, C.~W. Oates, and S.~A. Diddams,
  \enquote{{Generation of ultrastable microwaves via optical frequency
  division},} { {Nature Photonics}} \textbf{5}, 425--429
  (2011).

\bibitem{Wu2018}
J.~Wu, X.~Xu, T.~G. Nguyen, S.~T. Chu, B.~E. Little, R.~Morandotti,
  A.~Mitchell, and D.~J. Moss, \enquote{{RF Photonics: An Optical Microcombs'
  Perspective},} { {IEEE Journal of Selected Topics in
  Quantum Electronics}} \textbf{24}, 6101020 (2018).

\bibitem{Lucas2020}
E.~Lucas, P.~Brochard, R.~Bouchand, S.~Schilt, T.~S{\"{u}}dmeyer, and T.~J.
  Kippenberg, \enquote{{Ultralow-noise photonic microwave synthesis using a
  soliton microcomb-based transfer oscillator},} { {Nature
  Communications}} \textbf{11}, 1--8 (2020).

\bibitem{Li2008}
C.~H. Li, A.~J. Benedick, P.~Fendel, A.~G. Glenday, F.~X. K{\"{a}}rtner, D.~F.
  Phillips, D.~Sasselov, A.~Szentgyorgyi, and R.~L. Walsworth, \enquote{{A
  laser frequency comb that enables radial velocity measurements with a
  precision of 1 cm s-1},} { {Nature}} \textbf{452},
  610--612 (2008).

\bibitem{Metcalf2019}
A.~J. Metcalf, T.~Anderson, C.~F. Bender, S.~Blakeslee, W.~Brand, D.~R.
  Carlson, W.~D. Cochran, S.~A. Diddams, M.~Endl, C.~Fredrick, S.~Halverson,
  D.~D. Hickstein, F.~Hearty, J.~Jennings, S.~Kanodia, K.~F. Kaplan, E.~Levi,
  E.~Lubar, S.~Mahadevan, A.~Monson, J.~P. Ninan, C.~Nitroy, S.~Osterman, S.~B.
  Papp, F.~Quinlan, L.~Ramsey, P.~Robertson, A.~Roy, C.~Schwab, S.~Sigurdsson,
  K.~Srinivasan, G.~Stefansson, D.~A. Sterner, R.~Terrien, A.~Wolszczan, J.~T.
  Wright, and G.~Ycas, \enquote{{Stellar spectroscopy in the near-infrared with
  a laser frequency comb},} { {Optica}} \textbf{6},
  233--239 (2019).

\bibitem{Marin-Palomo2017}
P.~Marin-Palomo, J.~N. Kemal, M.~Karpov, A.~Kordts, J.~Pfeifle, M.~H. Pfeiffer,
  P.~Trocha, S.~Wolf, V.~Brasch, M.~H. Anderson, R.~Rosenberger, K.~Vijayan,
  W.~Freude, T.~J. Kippenberg, and C.~Koos, \enquote{{Microresonator-based
  solitons for massively parallel coherent optical communications},}
  { {Nature}} \textbf{546}, 274--279 (2017).

\bibitem{Fulop2018}
A.~F{\"{u}}l{\"{o}}p, M.~Mazur, A.~Lorences-Riesgo, {\'{O}}.~B. Helgason, P.~H.
  Wang, Y.~Xuan, D.~E. Leaird, M.~Qi, P.~A. Andrekson, A.~M. Weiner, and
  V.~Torres-Company, \enquote{{High-order coherent communications using
  mode-locked dark-pulse Kerr combs from microresonators},}
  { {Nature Communications}} \textbf{9}, 1--8 (2018).

\bibitem{Corcoran2020}
B.~Corcoran, M.~Tan, X.~Xu, A.~Boes, J.~Wu, T.~G. Nguyen, S.~T. Chu, B.~E.
  Little, R.~Morandotti, A.~Mitchell, and D.~J. Moss, \enquote{{Ultra-dense
  optical data transmission over standard fibre with a single chip source},}
  { {Nature Communications}} \textbf{11}, 1--7 (2020).

\bibitem{Li2018}
N.~Li, D.~Vermeulen, Z.~Su, E.~S. Magden, M.~Xin, N.~Singh, A.~Ruocco,
  J.~Notaros, C.~V. Poulton, E.~Timurdogan, C.~Baiocco, and M.~R. Watts,
  \enquote{{Monolithically integrated erbium-doped tunable laser on a
  CMOS-compatible silicon photonics platform},} { {Optics
  Express}} \textbf{26}, 16200--16211 (2018).

\bibitem{Shtyrkova2019}
K.~Shtyrkova, P.~T. Callahan, N.~Li, E.~S. Magden, A.~Ruocco, D.~Vermeulen,
  F.~X. K{\"{a}}rtner, M.~R. Watts, and E.~P. Ippen, \enquote{{Integrated
  CMOS-compatible Q-switched mode-locked lasers at 1900nm with an on-chip
  artificial saturable absorber},} { {Optics Express}}
  \textbf{27}, 3542--3556 (2019).

\bibitem{Huang2019}
D.~Huang, M.~A. Tran, J.~Guo, J.~Peters, T.~Komljenovic, A.~Malik, P.~A.
  Morton, and J.~E. Bowers, \enquote{{High-power sub-kHz linewidth lasers fully
  integrated on silicon},} { {Optica}} \textbf{6}, 745--752
  (2019).

\bibitem{Bhardwaj2020}
A.~Bhardwaj, R.~Bustos-Ramirez, G.~E. Hoefler, A.~Dentai, M.~E. Plascak,
  F.~Kish, P.~J. Delfyett, and M.~C. Wu, \enquote{{A Monolithically Integrated
  Racetrack Colliding-Pulse Mode-Locked Laser with Pulse-Picking Modulator},}
  { {IEEE Journal of Quantum Electronics}} \textbf{56},
  1--8 (2020).

\bibitem{Yu2020a}
Q.~Yu, J.~Gao, N.~Ye, B.~Chen, K.~Sun, L.~Xie, K.~Srinivasan, M.~Zervas,
  G.~Navickaite, M.~Geiselmann, and A.~Beling, \enquote{{Heterogeneous
  photodiodes on silicon nitride waveguides},} { {Optics
  Express}} \textbf{28}, 14824--14830 (2020).

\bibitem{Lu2019}
J.~Lu, J.~B. Surya, X.~Liu, A.~W. Bruch, Z.~Gong, Y.~Xu, and H.~X. Tang,
  \enquote{{Periodically poled thin-film lithium niobate microring resonators
  with a second-harmonic generation efficiency of 250,000{\%}/W},}
  { {Optica}} \textbf{6}, 1455--1460 (2019).

\bibitem{Bruch2020}
A.~W. Bruch, X.~Liu, Z.~Gong, J.~B. Surya, M.~Li, C.-L. Zou, and H.~X. Tang,
  \enquote{{Pockels Soliton Microcomb},}
  { {arXiv:2004.07708}}  (2020).

\bibitem{Hickstein2018}
D.~D. Hickstein, G.~C. Kerber, D.~R. Carlson, L.~Chang, D.~Westly,
  K.~Srinivasan, A.~Kowligy, J.~E. Bowers, S.~A. Diddams, and S.~B. Papp,
  \enquote{{Quasi-Phase-Matched Supercontinuum Generation in Photonic
  Waveguides},} { {Physical Review Letters}} \textbf{120},
  053903 (2018).

\bibitem{Singh2018}
N.~Singh, M.~Xin, D.~Vermeulen, K.~Shtyrkova, N.~Li, P.~T. Callahan, E.~S.
  Magden, A.~Ruocco, N.~Fahrenkopf, C.~Baiocco, B.~P. Kuo, S.~Radic, E.~Ippen,
  F.~X. K{\"{a}}rtner, and M.~R. Watts, \enquote{{Octave-spanning coherent
  supercontinuum generation in silicon on insulator from 1.06 $\mu$m to beyond
  2.4 $\mu$m},} { {Light: Science and Applications}}
  \textbf{7}, 17131 (2018).

\bibitem{Rao2019}
A.~Rao, K.~Abdelsalam, T.~Sjaardema, A.~Honardoost, G.~F. Camacho-Gonzalez, and
  S.~Fathpour, \enquote{{ Actively-monitored periodic-poling in thin-film
  lithium niobate photonic waveguides with ultrahigh nonlinear conversion
  efficiency of 4600 {\%}W$^{-1}$cm$^{-2}$ },} { {Optics
  Express}} \textbf{27}, 25920--25930 (2019).

\bibitem{Stanton2020}
E.~J. Stanton, J.~Chiles, N.~Nader, G.~Moody, N.~Volet, L.~Chang, J.~E. Bowers,
  S.~{Woo Nam}, and R.~P. Mirin, \enquote{{Efficient second harmonic generation
  in nanophotonic GaAs-on-insulator waveguides},} { {Optics
  Express}} \textbf{28}, 9521--9532 (2020).

\bibitem{Chang2017}
L.~Chang, M.~H.~P. Pfeiffer, N.~Volet, M.~Zervas, J.~D. Peters, C.~L.
  Manganelli, E.~J. Stanton, Y.~Li, T.~J. Kippenberg, and J.~E. Bowers,
  \enquote{{Heterogeneous integration of lithium niobate and silicon nitride
  waveguides for wafer-scale photonic integrated circuits on silicon},}
  { {Optics Letters}} \textbf{42}, 803--806 (2017).

\bibitem{Honardoost2018}
A.~Honardoost, G.~F.~C. Gonzalez, S.~Khan, M.~Malinowski, A.~Rao, J.~E.
  Tremblay, A.~Yadav, K.~Richardson, M.~C. Wu, and S.~Fathpour,
  \enquote{{Cascaded Integration of Optical Waveguides with Third-Order
  Nonlinearity with Lithium Niobate Waveguides on Silicon Substrates},}
  { {IEEE Photonics Journal}} \textbf{10}, 4500909 (2018).

\bibitem{Stanton2019}
E.~J. Stanton, L.~Chang, W.~Xie, A.~Malik, J.~Peters, J.~Chiles, N.~Nader,
  G.~Navickaite, D.~Sacchetto, M.~Zervas, K.~Srinivasan, J.~E. Bowers, S.~B.
  Papp, S.~W. Nam, and R.~P. Mirin, \enquote{{On-chip polarization rotator for
  type i second harmonic generation},} { {APL Photonics}}
  \textbf{4} (2019).

\bibitem{Stern2018}
B.~Stern, X.~Ji, Y.~Okawachi, A.~L. Gaeta, and M.~Lipson,
  \enquote{{Battery-operated integrated frequency comb generator},}
  { {Nature}} \textbf{562}, 401--405 (2018).

\bibitem{Raja2019}
A.~S. Raja, A.~S. Voloshin, H.~Guo, S.~E. Agafonova, J.~Liu, A.~S.
  Gorodnitskiy, M.~Karpov, N.~G. Pavlov, E.~Lucas, R.~R. Galiev, A.~E.
  Shitikov, J.~D. Jost, M.~L. Gorodetsky, and T.~J. Kippenberg,
  \enquote{{Electrically pumped photonic integrated soliton microcomb},}
  { {Nature Communications}} \textbf{10}, 680 (2019).

\bibitem{Shen2020}
B.~Shen, L.~Chang, J.~Liu, H.~Wang, Q.~F. Yang, C.~Xiang, R.~N. Wang, J.~He,
  T.~Liu, W.~Xie, J.~Guo, D.~Kinghorn, L.~Wu, Q.~X. Ji, T.~J. Kippenberg,
  K.~Vahala, and J.~E. Bowers, \enquote{{Integrated turnkey soliton
  microcombs},} { {Nature}} \textbf{582}, 365--369 (2020).

\bibitem{Hall2006}
J.~L. Hall, \enquote{{Nobel lecture: Defining and measuring optical
  frequencies},} { {Reviews of Modern Physics}}
  \textbf{78}, 1279--1295 (2006).

\bibitem{Huang2016}
S.~W. Huang, J.~Yang, M.~Yu, B.~H. McGuyer, D.~L. Kwong, T.~Zelevinsky, and
  C.~W. Wong, \enquote{{A broadband chip-scale optical frequency synthesizer at
  2.7 x 10$^{-16}$ relative uncertainty},} { {Science
  Advances}} \textbf{2}, e1501489 (2016).

\bibitem{Arafin2017a}
S.~Arafin, A.~Simsek, S.~K. Kim, W.~Liang, D.~Eliyahu, G.~Morrison,
  M.~Mashanovitch, A.~Matsko, L.~Johansson, L.~Maleki, M.~J. Rodwell, and L.~A.
  Coldren, \enquote{{Power-Efficient Kerr Frequency Comb Based Tunable Optical
  Source},} { {IEEE Photonics Journal}} \textbf{9} (2017).

\bibitem{Arafin2017}
S.~Arafin, A.~Simsek, S.-K. Kim, S.~Dwivedi, W.~Liang, D.~Eliyahu, J.~Klamkin,
  A.~Matsko, L.~Johansson, L.~Maleki, M.~Rodwell, and L.~Coldren,
  \enquote{{Towards chip-scale optical frequency synthesis based on optical
  heterodyne phase-locked loop},} { {Optics Express}}
  \textbf{25}, 681--695 (2017).

\bibitem{Xin2019}
M.~Xin, N.~Li, N.~Singh, A.~Ruocco, Z.~Su, E.~S. Magden, J.~Notaros,
  D.~Vermeulen, E.~P. Ippen, M.~R. Watts, and F.~X. K{\"{a}}rtner,
  \enquote{{Optical frequency synthesizer with an integrated erbium tunable
  laser},} { {Light: Science and Applications}} \textbf{8},
  1--8 (2019).

\bibitem{Singh2020}
N.~Singh, M.~Xin, N.~Li, D.~Vermeulen, A.~Ruocco, E.~S. Magden, K.~Shtyrkova,
  E.~Ippen, F.~X. K{\"{a}}rtner, and M.~R. Watts, \enquote{{Silicon Photonics
  Optical Frequency Synthesizer},} { {Laser and Photonics
  Reviews}} \textbf{14}, 1900449 (2020).

\bibitem{Costanzo2020}
R.~Costanzo and S.~M. Bowers, \enquote{{A 10-GHz Bandwidth Transimpedance
  Amplifier with Input DC Photocurrent Compensation Loop},}
  { {IEEE Microwave and Wireless Components Letters}}
  \textbf{30}, 673--676 (2020).

\bibitem{Kleijn2014}
E.~Kleijn, D.~Melati, A.~Melloni, T.~{De Vries}, M.~K. Smit, and X.~J.
  Leijtens, \enquote{{Multimode interference couplers with reduced parasitic
  reflections},} { {IEEE Photonics Technology Letters}}
  \textbf{26}, 408--410 (2014).

\bibitem{Manolatou1999}
C.~Manolatou, M.~J. Khan, S.~Fan, P.~R. Villeneuve, H.~A. Haus, and J.~D.
  Joannopoulos, \enquote{{Coupling of modes analysis of resonant channel
  add-drop filters},} { {IEEE Journal of Quantum
  Electronics}} \textbf{35}, 1322--1331 (1999).

\bibitem{Moille2019}
G.~Moille, Q.~Li, T.~C. Briles, S.-P. Yu, T.~Drake, X.~Lu, A.~Rao, D.~Westly,
  S.~B. Papp, and K.~Srinivasan, \enquote{{Broadband resonator-waveguide
  coupling for efficient extraction of octave-spanning microcombs},}
  { {Optics Letters}} \textbf{44}, 4737--4740 (2019).

\bibitem{Barwicz2004}
T.~Barwicz, M.~A. Popovic, P.~T. Rakich, M.~R. Watts, H.~A. Haus, E.~P. Ippen,
  and H.~I. Smith, \enquote{{Microring-resonator-based add-drop filters in SiN:
  fabrication and analysis},} { {Optics Express}}
  \textbf{12}, 1437 (2004).

\bibitem{Amatya2008}
R.~Amatya, C.~W. Holzwarth, H.~I. Smith, and R.~J. Ram, \enquote{{Precision
  tunable silicon compatible microring filters},} { {IEEE
  Photonics Technology Letters}} \textbf{20}, 1739--1741 (2008).

\end{thebibliography}

\noindent \textbf{Acknowledgements} The authors acknowledge funding from the DARPA DODOS and NIST-on-a-chip programs. A.R. and X.L. acknowledge support under the Cooperative Research Agreement between the University of Maryland and NIST-CNST, Award no. 70NANB10H193.\\

\noindent \textbf{Additional Information} Correspondence and requests for materials should be addressed to A.R. and K.S. \\

\noindent \textbf{Competing financial interests} The authors declare no competing financial interests.

\end{document}

\title{Integrated photonic interposers for processing octave-spanning microresonator frequency combs: \\* 
Supplementary Information
}

\author{Ashutosh Rao}\email{ashutosh.rao@nist.gov}
\affiliation{Microsystems and Nanotechnology Division, Physical Measurement Laboratory, National Institute of Standards and Technology, Gaithersburg, MD 20899, USA}\affiliation{Maryland NanoCenter, University of Maryland,
College Park, MD 20742, USA}

\author{Gregory Moille}
\affiliation{Microsystems and Nanotechnology Division, Physical Measurement Laboratory, National Institute of Standards and Technology, Gaithersburg, MD 20899, USA}\affiliation{Joint Quantum Institute, NIST/University of Maryland, College Park, MD 20742, USA}
\author{Xiyuan Lu}
\affiliation{Microsystems and Nanotechnology Division, Physical Measurement Laboratory, National Institute of Standards and Technology, Gaithersburg, MD 20899, USA}\affiliation{Maryland NanoCenter, University of Maryland,
College Park, MD 20742, USA}
\author{Daron A. Westly}
\affiliation{Microsystems and Nanotechnology Division, Physical Measurement Laboratory, National Institute of Standards and Technology, Gaithersburg, MD 20899, USA}

\author{Davide Sacchetto}
\affiliation{Ligentec, EPFL Innovation Park, Bâtiment C, Lausanne, Switzerland}
\author{Michael Geiselmann}
\affiliation{Ligentec, EPFL Innovation Park, Bâtiment C, Lausanne, Switzerland}
\author{Michael Zervas}
\affiliation{Ligentec, EPFL Innovation Park, Bâtiment C, Lausanne, Switzerland}

\author{Scott B. Papp}
\affiliation{Time and Frequency Division, Physical Measurement Laboratory, National Institute of Standards and Technology, Boulder, CO 80305, USA}
\affiliation{Department of Physics, University of Colorado, Boulder, CO 80309, USA}
\author{John Bowers}
\affiliation{Department of Electrical and Computer Engineering, University of California, Santa Barbara, CA 93106, USA}

\author{Kartik Srinivasan} \email{kartik.srinivasan@nist.gov}
\affiliation{Microsystems and Nanotechnology Division, Physical Measurement Laboratory, National Institute of Standards and Technology, Gaithersburg, MD 20899, USA}\affiliation{Joint Quantum Institute, NIST/University of Maryland, College Park, MD 20742, USA}

\date{\today}

\begin{abstract}
\noindent This document contains  additional details of device fabrication, device design, and experimental setups.
\end{abstract}

\maketitle

\begin{center}
\begin{figure*}
\begin{center}
\includegraphics[width=1\linewidth]{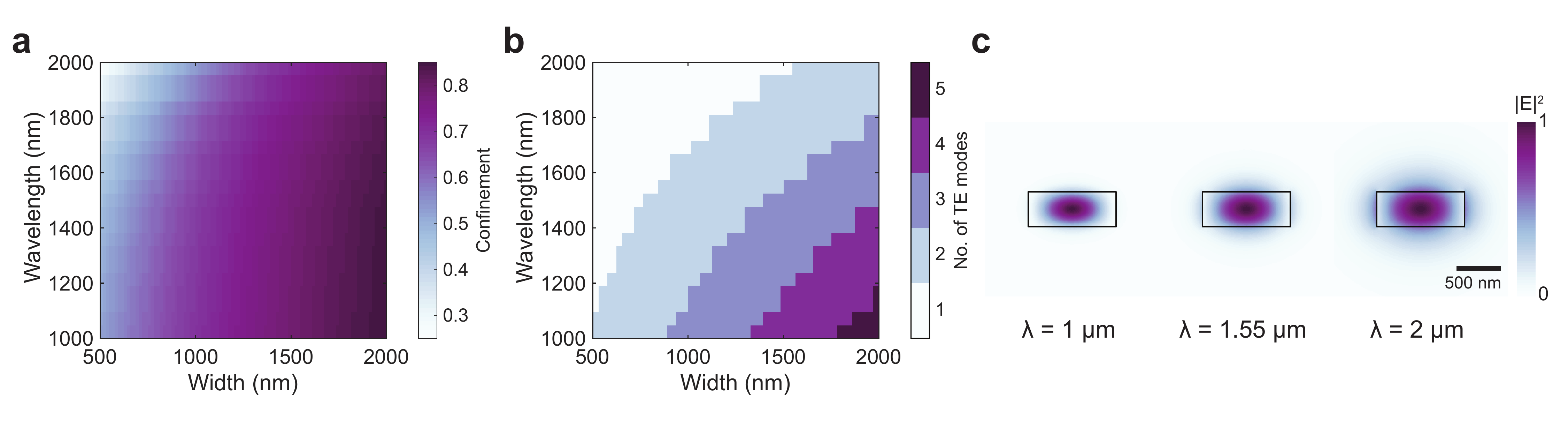}
\caption{\textbf{Photonics Platform}. 
\textbf{a},\textbf{b}. Optical confinement and number of modes for channel waveguides in a 400~nm thick silicon nitride film with silicon dioxide upper and lower cladding. The optical confinement here is defined as $\eta = \iint_{core} \lvert \textbf{E}(x,y)\rvert^2 \,dx\,dy/ \iint\lvert \textbf{E}(x,y)\rvert^2 \,dx\,dy $. A nominal waveguide width of 1 ${\rm \mu}$m balances the confinement and number of modes across the octave, and is followed by additional tapering throughout the interposer to reach the target dimensions of specific elements (e.g., the dichroics). \textbf{c}. Waveguide transverse electric field modes simulated for wavelengths of 1~${\rm \mu}$m, 1.55~${\rm \mu}$m, and 2~${\rm \mu}$m, for a waveguide width of 1~${\rm \mu}$m. In keeping with \textbf{b}, $ \lvert \textbf{E}(x,y)\rvert^2$ is plotted.}
\label{fig:FigS1_new}
\end{center}
\end{figure*}
\end{center}

\begin{center}
\begin{figure*}
\begin{center}
\includegraphics[width=1\linewidth]{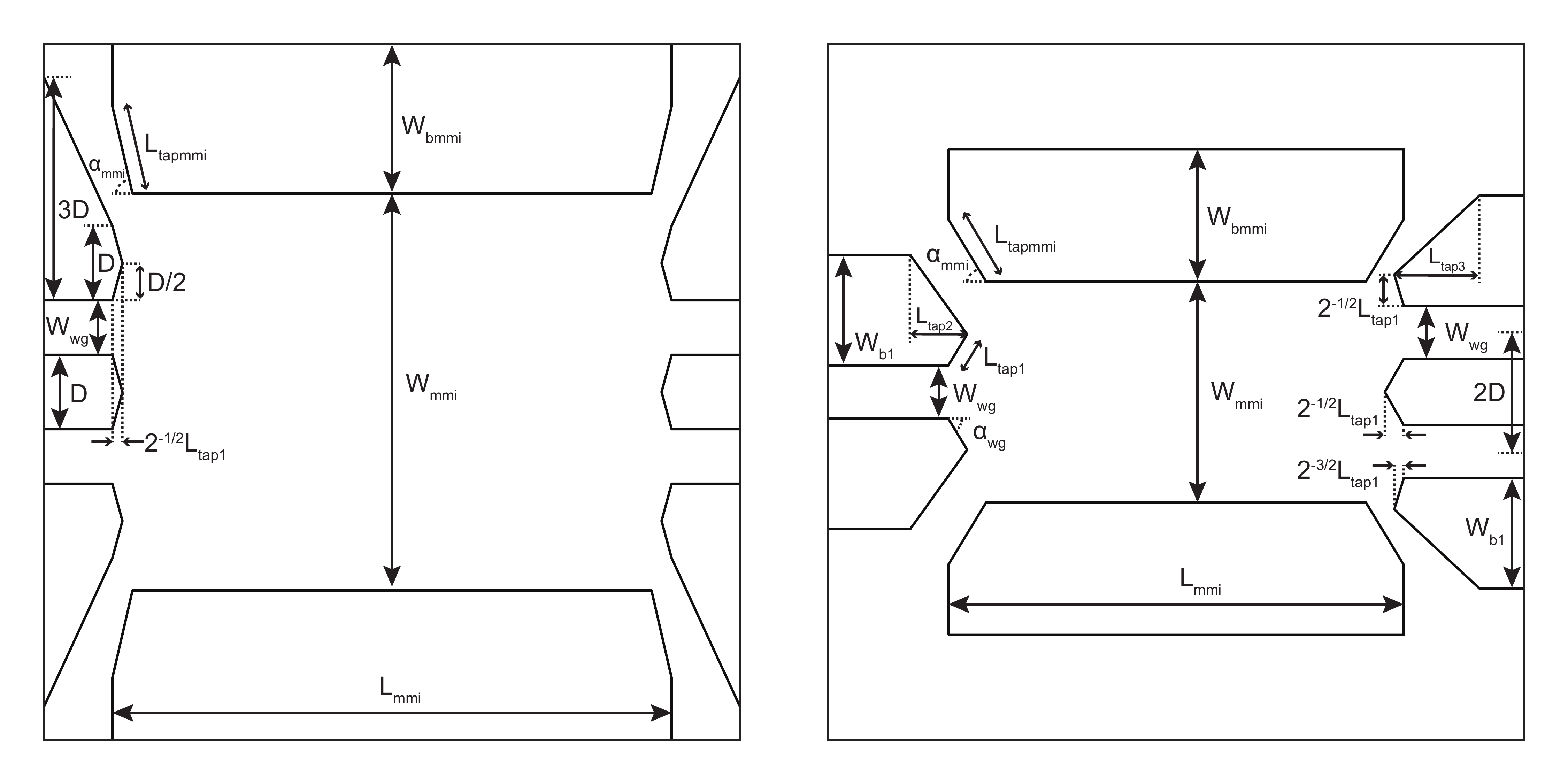}
\caption{\textbf{Multimode Interferometers}. Detailed design schematics for 2x2 and 1x2 multimode interferometers.}
\label{fig:FigS2}
\end{center}
\end{figure*}
\end{center}

\section{Device Fabrication}

All devices used here are fabricated on silicon dioxide (SiO$_2$)-clad silicon nitride (${\rm Si_3N_4}$) photonic platforms. Low pressure chemical vapor deposition is used to deposit these ${\rm Si_3N_4}$ layers. The Nanolithography Toolbox~\cite{Balram2016}, a free package developed by the NIST Center for Nanoscale Science and Technology, was used for all device layouts. Broadband ellipsometry was used along with an extended Sellmeier model to evaluate the refractive index across the wavelength range of interest. All devices are fabricated on 100 mm silicon wafers. The octave spanning microcomb, the interposer elements (multimode interferometers, ring filters, and dichroics), and the bilayer microcomb are fabricated at Ligentec using deep-UV lithography. All of these are patterened via reactive ion etching, except for the microcomb layer of the bilayer microcomb, which is patterned using a damascene process. The integrated microcomb spectral filter is fabricated at NIST using electron-beam lithography and reactive ion etching.

\section{Device designs and measurements}

\subsection{Photonics Platform}

Figures~S1a and S1b show the variation of modal confinement of the fundamental transverse-electric (TE) mode and the number of TE modes with wavelength and waveguide width. Figure~S1c shows simulated TE eigenmodes for a waveguide width of 1~${\rm \mu}$m that nominally balances these criteria.

\subsection{Multimode Interferometers}

Figure~S2 and Table~S1 show the optimized design parameters of the multimode interferometers. Initial designs for a standard geometry~\cite{Soldano1995} were adapted and optimized for the butterfly geometry~\cite{Kleijn2014} used here through 3D finite difference time domain (FDTD) simulations.

\begin{center}
\begin{table}
\begin{center}
 \begin{tabular}{||c| c| c| c||} 
 \hline
 Parameter & 2x2 1050~nm & 2x2 1550~nm & 1x2 1550~nm \\ [0.5ex] 
 \hline\hline
 $L_{mmi}$ ($\mu$m) & 23 & 49 & 15 \\ 
 \hline
 $W_{mmi}$ ($\mu$m) & 7 & 8 & 5 \\
 \hline
 $W_{bmmi}$ ($\mu$m) & 3 & 3 & 3 \\
 \hline
 $L_{tapmmi}$ & 2$L_{tap1}$ & 2$L_{tap1}$ & 2 \\
 \hline
 $\alpha_{mmi}$ & 45$^{\circ}$ & 45$^{\circ}$ & 45$^{\circ}$ \\  
 \hline
 $D$ ($\mu$m) & 0.75 & 1.5 & 1.35 \\  
 \hline
 $W_{wg}$ ($\mu$m) & 1 & 1.1 & 1.2 \\  
 \hline
 $\alpha_{wg}$ & - & - & 45$^{\circ}$ \\  
 \hline
 $W_{b1}$ ($\mu$m) & - & - & 2.5 \\  
 \hline
 $L_{tap1}$ ($\mu$m) & 1.25 & 1.25 & 1 \\  
 \hline
 $L_{tap2}$ & - & - & 3$L_{tap1}/\sqrt{2}$ \\  
 \hline
 $L_{tap3}$ & - & - & 4.5$L_{tap1}/\sqrt{2}$ \\ [1ex]  
 \hline
\end{tabular}
\end{center}
\caption{\label{tab:TableS1} Geometrical parameters for multimode interferometers.}
\end{table}
\end{center}

\subsection{Ring Filters}

The filter response depends on the intrinsic and coupling Q, as discussed in the main text. Figure~S3 shows the variation of coupling Q with the coupling gap between the microring and bus waveguide, calculated using coupled mode theory~\cite{Moille2019}. The corresponding parameters used  are ring radius = 50~$\mu$m, ring width = 50~$\mu$m, and bus waveguide width = 1~$\mu$m.

\begin{center}
\begin{figure*}
\begin{center}
\includegraphics[width=1\linewidth]{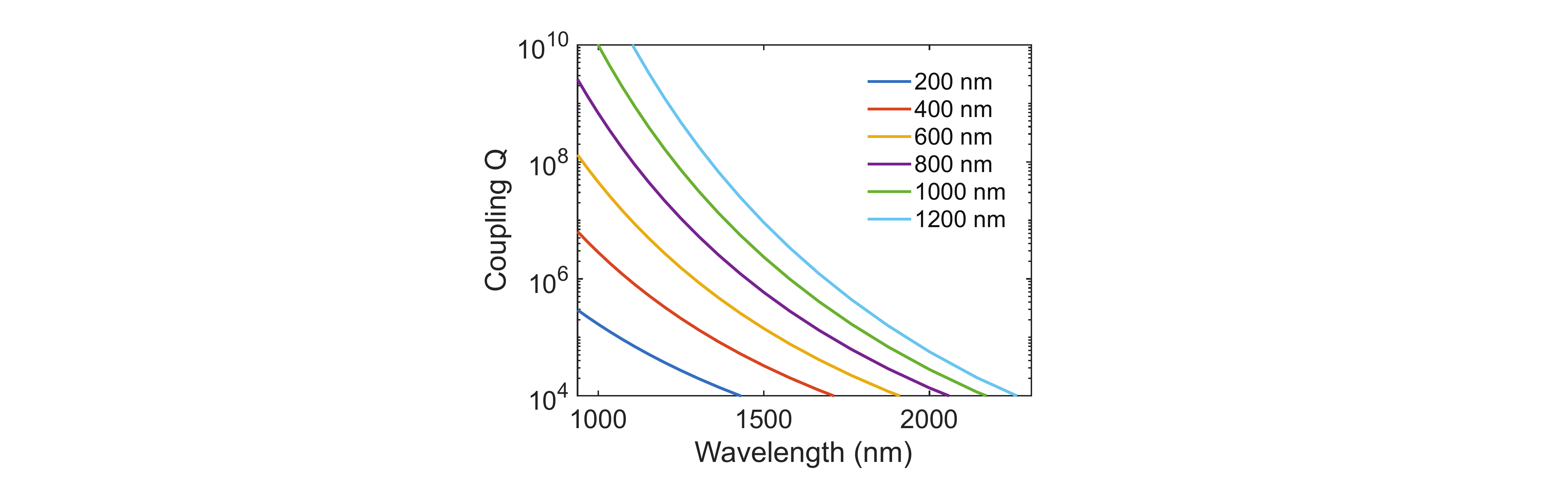}
\caption{\textbf{Ring filter coupling} Simultaed variation of ring filter coupling Q with coupling gap for a straight bus waveguide. The ring filter is meant to operate around 1550~nm wavelength. For a coupling Q $\approx 10^5$ at 1550 nm, the filter is severely undercoupled at 1000~nm wavelength, as desired, with coupling Q $\approx 5\times10^7$.}
\label{fig:FigS3}
\end{center}
\end{figure*}
\end{center}

\subsection{Dichroics}

\begin{center}
\begin{figure*}
\begin{center}
\includegraphics[width=1\linewidth]{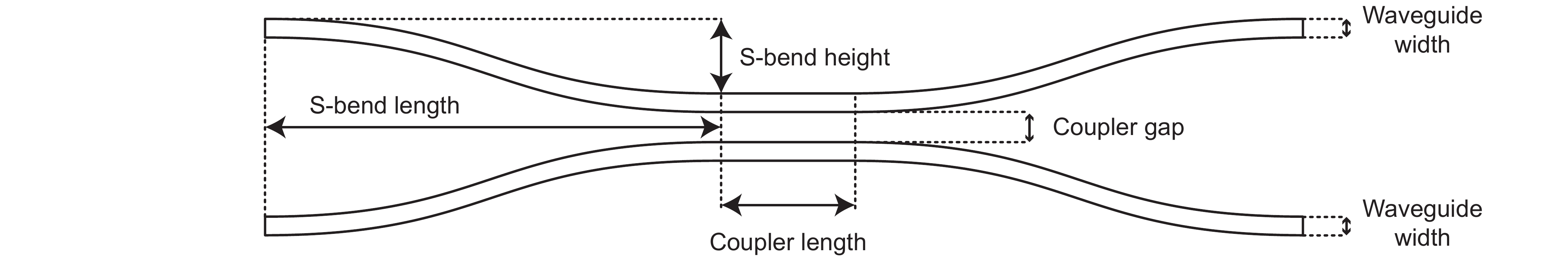}
\caption{\textbf{Dichroic couplers} Schematic of dichroic couplers. Dichroic 1 filters out 2 $\mu$m light into its cross port, and dichroic 2 filters out 1.55 $\mu$m light into its cross port.}
\label{fig:FigS4}
\end{center}
\end{figure*}
\end{center}

\begin{center}
\begin{table}
\begin{center}
 \begin{tabular}{||c| c| c| c||} 
 \hline
 Parameter & Dichroic 1 & Dichroic 2 \\ [0.5ex] 
 \hline\hline
 Coupler length ($\mu$m) & 50 & 170 \\ 
 \hline
 Coupling gap ($\mu$m) & 1.25 & 2.5  \\
 \hline
 Waveguide width ($\mu$m) & 0.5   & 0.7  \\
 \hline
 S-bend length ($\mu$m) & 100 & 100 \\
 \hline
 S-bend height ($\mu$m) & 12.5 & 12.5  \\ [1ex] 
 \hline
\end{tabular}
\end{center}
\caption{\label{tab:TableS2} Geometrical parameters (in ${\rm \mu}$m) for dichroic couplers. Dichroic 1 filters out 2 $\mu$m light, and dichroic 2 filters out 1.55 $\mu$m light.}
\end{table}
\end{center}

\begin{center}
\begin{figure*}
\begin{center}
\includegraphics[width=1\linewidth]{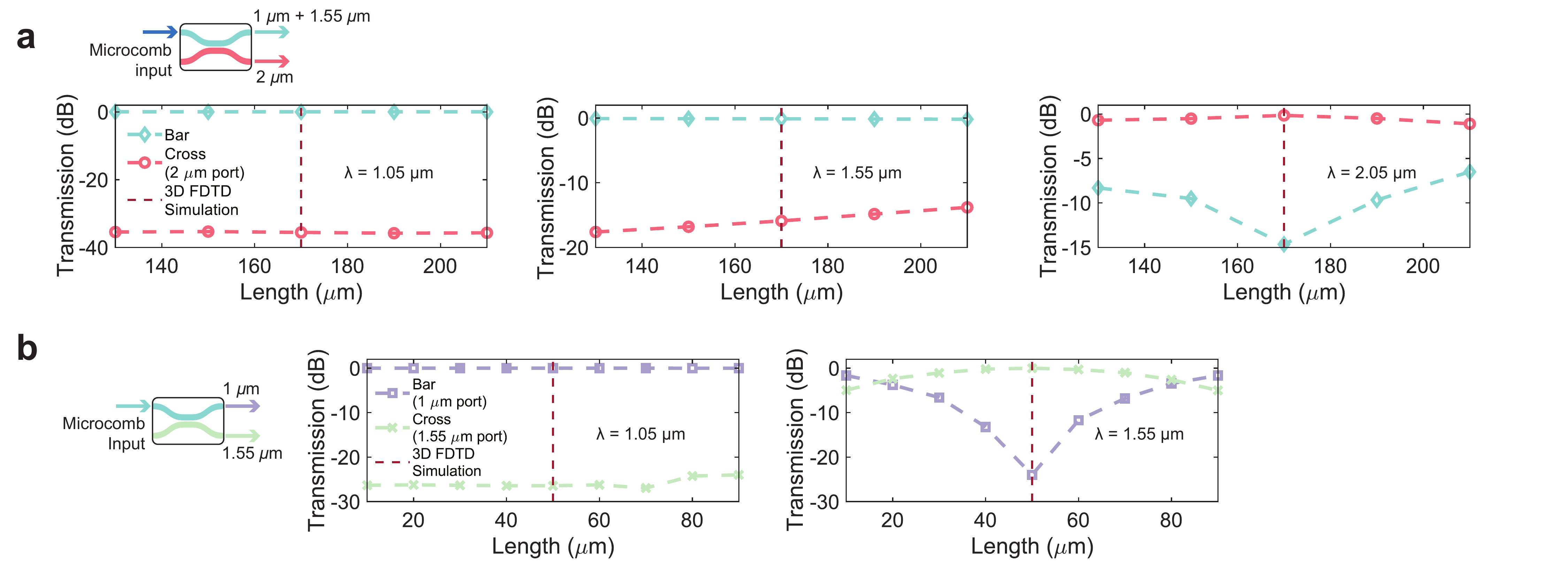}
\caption{\textbf{Continuous-wave measurements of dichroic couplers} \textbf{a}. First dichroic, whose purpose is to separate 2~$\mu$m light from shorter wavelengths. \textbf{b}. Second dichroic, whose purpose is to separate 1.55~$\mu$m light from shorter wavelengths. Both dichroics offer optimal performance for coupling lengths that are in agreement with optimized FDTD simulations. The uncertainties in excess loss corresponding to one standard deviation in transmission are less than 0.25~dB at the three wavelengths.}
\label{fig:FigS5}
\end{center}
\end{figure*}
\end{center}

The dichroic couplers (schematic shown in Fig.~S4) used here are based on the strong dispersion, across the octave bandwidth, of the evanescent decay of the optical mode outside the waveguide core. Qualitative starting points for waveguide widths can be found in Fig.~S1, which shows the optical confinement and is therefore indicative of the evanescent decay of the fundamental TE modes. Quantitatively, initial device parameters such as waveguide width and coupling gap are determined through finite-element-method based eigenmode simulations of the supermodes of uniform couplers. The nominal coupling lengths extracted from these supermode simulations are used as starting points for 3D FDTD simulations that take into account S-bends at the input and output of the dichroics. Table~S2 shows the design parameters of the two optimized dichroics. Figure~S5 shows the variation of dichroic coupler performance, extracted from continuous-wave measurements at wavelengths of 1.05~${\rm \mu}$m, 1.55~${\rm \mu}$m, and 2.05~${\rm \mu}$m. with coupling lengths, with optimal performance measured for the optimized designs.

\subsection{Broadband Bilayer taper}

\begin{center}
\begin{figure*}
\begin{center}
\includegraphics[width=1\linewidth]{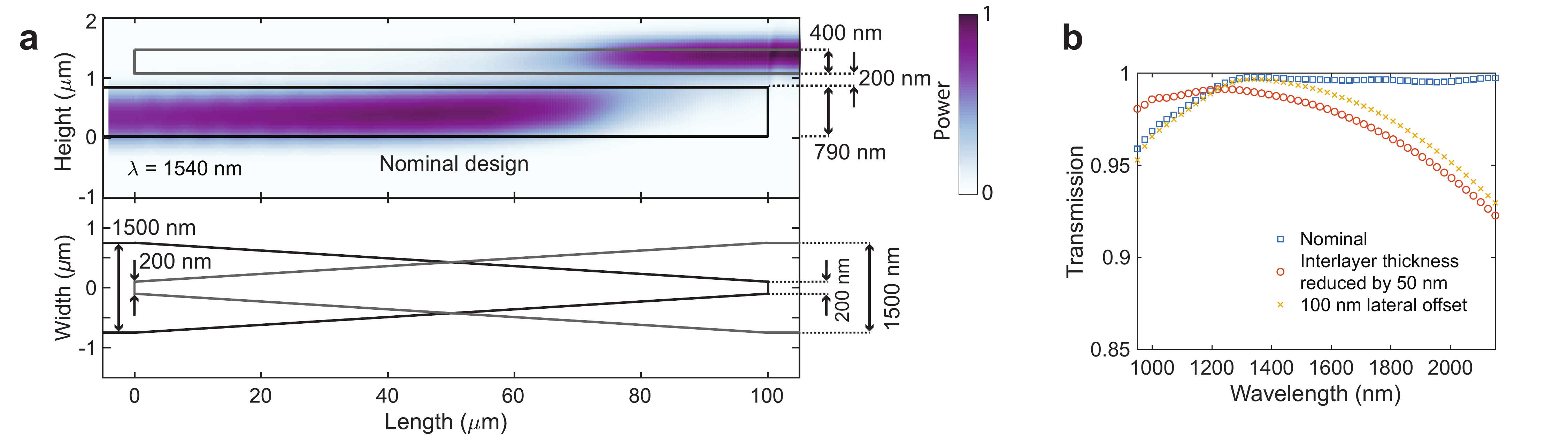}
\caption{\textbf{Bilayer coupler} \textbf{a}. Cross section and top views of the bilayer coupler. Both layers are linearly tapered from 200~nm to 1500~nm. \textbf{b}. Transmission spectra for a reduced SiO$_2$ interlayer thickness of 150~nm, and a 100~nm lateral misalignment between the tapers, compared to the nominal design.}
\label{fig:FigS6}
\end{center}
\end{figure*}
\end{center}

Figure~S6a shows a detailed schematic of the broadband bilayer taper. The transfer of light here requires a balance of the phase-matching behind the bilayer coupling across the octave bandwidth. We limit the minimum widths of the tapers in accordance with the corresponding fabrication process (deep-UV lithography), and a broadband 3D FDTD sweep is used to determine the overall taper length. For a taper shorter than the optimal 100~${\rm \mu}$m, the bilayer coupling is reduced for shorter wavelengths close to 1~${\rm \mu}$m. Figure~S6b shows the tolerance in taper transmission to interlayer thickness and taper misalignment.

\begin{center}
\begin{figure*}
\begin{center}
\includegraphics[width=1\linewidth]{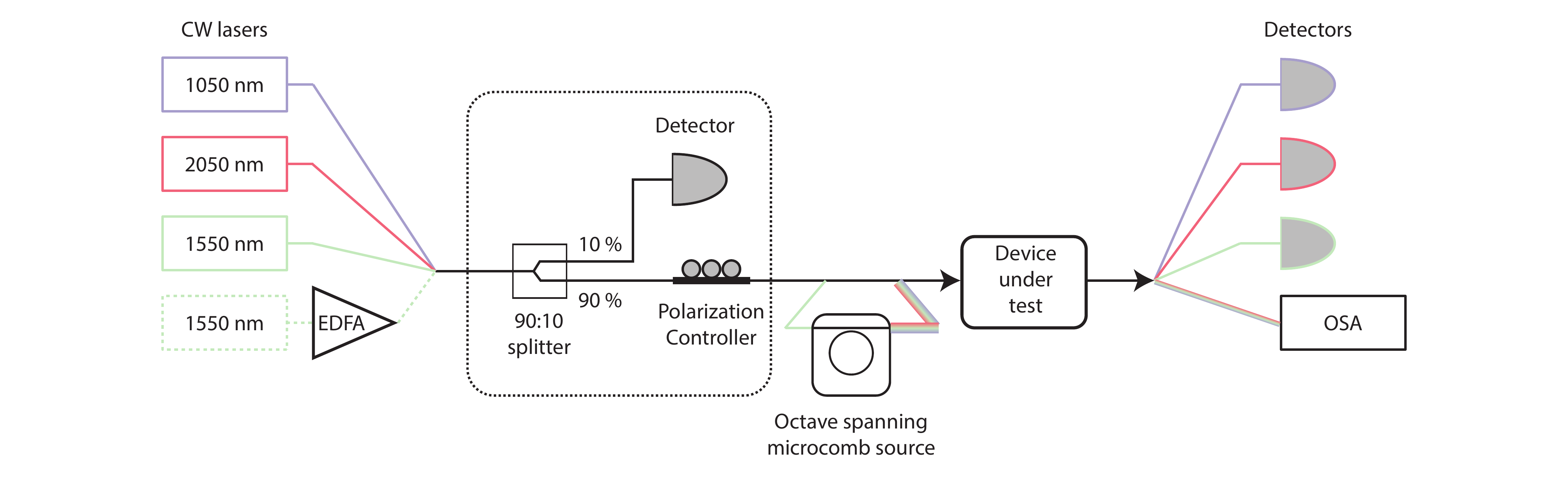}
\caption{\textbf{Experimental setups} Multiple experimental setups are used throughout this work, depending on the combination of the device under test and the corresponding inputs. The devices tested are multimode interferometers, ring filters, dichroics, integrated spectral microcomb filter, and the bilayer microcomb. The inputs used are continuous wave lasers at 1050~nm, 1550~nm, and 2050~nm, and an octave spanning microcomb pumped around 1550~nm. Polarization maintaining fiber is used to couple the octave spanning microcomb to the corresponding devices under test. Two OSAs are used to cover the octave bandwidth of the microcomb. EDFA = Erbium-Doped Fiber Amplifier. OSA = Optical Spectrum Analyzer.}
\label{fig:FigS7}
\end{center}
\end{figure*}
\end{center}

\section{Experimental Setup}

The experimental setups used are shown in Fig.~S7, illustrating the different configurations used for measurements of the multimode interferometers, ring filters, dichroics, integrated spectral microcomb filter, and bilayer microcomb. Each continuous wave laser requires separate fiber components such as the 90:10 splitter and polarization controller, to satisfy the single mode criterion in the fiber. The detector following the 10 $\%$ port is used to assist in stabilizing the coupling to the device under test. TE polarization is used throughout all the measurements. Lensed optical fibers with focused spot sizes of $\approx$ 2.5~${\rm \mu}$m are used to couple light on and off the chips, aided by inverse tapers on the chips to match the mode profiles between the lensed fibers and waveguides.  

%\bibliographystyle{osajnl}
%\bibliography{References_interposer}